\documentclass[aps,prx,reprint,twocolumn,superscriptaddress,floatfix,nofootinbib,longbibliography]{revtex4-1}
\usepackage{epsfig,amsmath,amssymb,color,comment,physics}
\usepackage{placeins}
\usepackage[makeroom]{cancel}
\usepackage[caption=false]{subfig}
\usepackage{float}
\usepackage[countmax]{subfloat}
\usepackage[normalem]{ulem}
\usepackage[english]{babel}
\usepackage{dsfont}
\usepackage{braket}
\usepackage[bookmarks=true,colorlinks,linkcolor=OrangeRed,urlcolor=NavyBlue,citecolor=RoyalBlue]{hyperref}
\usepackage[dvipsnames]{xcolor}
\usepackage{graphicx}
\usepackage{cleveref}
\graphicspath{{./Figures/}}
\usepackage{amsmath, amssymb,latexsym,amsfonts}
\usepackage{orcidlink}
\usetikzlibrary{patterns}
\usepackage{bm}
\usepackage{hyperref}
\usepackage{subcaption}
\usepackage{quantikz}
\usepackage{tikz}
\usepackage{hhline}
\usepackage{moresize}
\usepackage{multirow}

\usepackage{tabularray}
\usepackage[export]{adjustbox}

\newcommand{\rb}{\rangle} 
\newcommand{\lb}{\langle} 
\newcommand{\h}{\hat} 
\newcommand{\X}{\hat{X}} 
\newcommand{\Y}{\hat{Y}} 
\newcommand{\Z}{\hat{Z}} 

\begin{document}

\preprint{APS/123-QED}

\title{Counterdiabatic quantum optimization for efficient state preparation in the Schwinger model}

\author{Ethan Laval${}^{\orcidlink{0009-0008-8592-8129}}$}
\email{E.B.Laval@soton.ac.uk}
\affiliation{School of Physics and Astronomy, University of Southampton, University Road, Southampton SO17 1BJ, UK}

\author{Bipasha Chakraborty${}^{\orcidlink{0000-0001-6667-329X}}$}
\email{B.Chakraborty@soton.ac.uk}
\affiliation{School of Physics and Astronomy, University of Southampton, University Road, Southampton SO17 1BJ, UK}

\author{Stefano Cipolla${}^{\orcidlink{0000-0002-8000-4719}}$}
\email{S.Cipolla@soton.ac.uk}
\affiliation{School of Mathematical Sciences, University of Southampton, University Road, Southampton SO17 1BJ, UK}

\date{\today}

\begin{abstract}

The Quantum Approximate Optimization Algorithm (QAOA), whose design is underpinned by the adiabatic theorem, is one of the leading variational quantum algorithms for preparing the ground states of gauge theories. A recently proposed variant, DC-QAOA, incorporates counterdiabatic driving to accelerate the adiabatic process, reducing both circuit depth and runtime. In this work, we present a systematic analysis, from both theoretical and computational perspectives, of DC-QAOA and related counterdiabatic variants for preparing the ground state of the Schwinger model, the paradigmatic (1+1)-dimensional lattice gauge theory. Benchmarking these methods against standard QAOA, we show that counterdiabatic driving improves ground-state preparation while lowering circuit depth requirements. Our results establish counterdiabatic protocols as a practical route towards efficient quantum simulation of gauge theories.

\end{abstract}

\maketitle


\section{Introduction}

The fundamental interactions of nature are described in terms of gauge theories. Quantum chromodynamics, the gauge theory that describes the strong interaction between quarks and gluons, is asymptotically free meaning that perturbation theory cannot be used when describing the theory at low energies. This motives the study of lattice gauge theories \cite{Munster_2000} which allow for non-perturbative calculations of gauge theories by discretising spacetime onto a lattice and evaluating numerical integrations of path integrals. However, it becomes a problem when the integrand of these integrals are non-real positive and highly oscillatory as then positive and negative contributions to the integral nearly cancel each other, requiring the integration to be done with very high precision. This is known as the sign problem \cite{Troyer_2005}, and it occurs in many situations, such as when there are chemical potentials \cite{Gattringer_2016} or topological terms \cite{Unsal_2012}.\\

A possible solution to this problem is to switch from a path integral formulation to a Hamiltonian formulation where there is no sign problem. However, it quickly becomes infeasible to simulate these Hamiltonians on classical computers due to the Hilbert space increasing exponentially with the number of lattice sites. Instead we look towards quantum computers which are more promising in this regard due to qubits being better suited to represent these large Hilbert spaces.\\ 

Recently, there has been a lot of research dedicated towards simulating lattice gauge theories on quantum technologies \cite{Ba_uls_2020,Funcke_2023}. A common simple gauge theory model used to develop and test these quantum simulations is the Schwinger model \cite{Martinez_2016,Muschik_2017,Klco_2018}. Performing quantum simulations of this theory involves discretising the theory on a lattice and reformulating the resulting Hamiltonian so that it acts on qubits. These simulations represent a first but foundational step towards the simulation of richer gauge theories, with quantum chromodynamics as the long-term goal. A common resource intensive task faced by these simulations is to prepare the ground state of the model's Hamiltonian using quantum algorithms. In quantum field theories, ground states of Hamiltonians correspond to vacuum states, which are fundamental to understanding the theory as it allows one to calculate excited states, vacuum expectation values, and scattering amplitudes.\\

The current state of quantum computing hardware is referred to as the noisy intermediate-scale quantum (NISQ) era \cite{Preskill_2018}, where physical quantum computers can have access to hundreds of qubits but are prone to noise and decoherence. This prevents them from being fault tolerant, and limits the number of gates and depths of the circuits they can run. Because of this, it is important to consider the resource cost of any algorithms we choose to run, and try to find algorithms with the lowest cost. For example, we can prepare the Schwinger model ground state using an approximate adiabatic evolution \cite{chakraborty2022}, but doing so requires a large amount of gates which makes it infeasible to run on a NISQ device.\\

A potential way to reduce the resource cost is to use variational quantum algorithms \cite{Cerezo_2021} which utilise both quantum and classical computation to reduce the resource cost of the quantum part of the algorithm. The classical part of the algorithm involves using a classical optimizer to train a parametrized quantum circuit, while the quantum part involves running and taking measurements of the circuit. These kinds of algorithms have previously been used to prepare the ground state of the Schwinger model \cite{Kokail_2019,funcke2022,li2024,fujii2025,Gustafson_2025}.\\

One of the leading variational quantum algorithms is the Quantum Approximate Optimization Algorithm (QAOA) \cite{farhi2014}, with this too having been applied to the Schwinger model \cite{bazavov2024,Tomlinson_2025}. The design of QAOA is underpinned by the adiabatic theorem which requires slow evolution times. This translates to large circuit depths in quantum algorithms. There have been a variety of techniques proposed, called shortcuts to adiabaticity \cite{Gu_ry_Odelin_2019}, which aim to speed up the adiabatic evolution process, with one of most common being counterdiabatic (CD) driving \cite{Demirplak_2003,Demirplak_2005,Berry_2009}. Recently, there has been a proposed variation of QAOA called Digitized Counterdiabatic QAOA (DC-QAOA) \cite{Chandarana_2022} that incorporates CD driving in order to get improved results at a reduced circuit depth, and with a lower number of variational circuit parameters. DC-QAOA has previously been applied to optimization problems such as MaxCut \cite{Chandarana_2022}, portfolio optimization \cite{Hegade_2022}, protein folding \cite{Chandarana_2023}, modular docking \cite{Ding_2024}, logistics scheduling \cite{Dalal_2024}, and bin packing \cite{Xu_2025}, with improved results over QAOA. There has also been proposed DC-QAOA variations, called CD-inspired \cite{Chandarana_2023} and CD-mixer \cite{Xu_2025}, that simplify the DC-QAOA circuit to further reduce the circuit depth.\\


In this paper, we will look at using QAOA and its CD variants to prepare the ground state of the Schwinger model and compare the performances and the final circuit resource cost of the algorithms. This is new type of problem to apply CD algorithms to, with them having mainly been applied to optimization problems. By doing so, we hope to find improved algorithms to prepare the ground state of the Schwinger model and gauge theories in general. We also provide a new method to choose the CD terms that appear in the CD algorithms which we call CD term fixing which will hopefully improve the performance of CD algorithms in the future. Additionally, we provide a new DC-QAOA variant which we call CD-prob.\\

This paper is structured as follows. In section \ref{Sec_2_background} we will go over the background of this paper, which includes describing in detail the algorithms QAOA, DC-QAOA, and the DC-QAOA variations CD-inspired, CD-mixer and CD-prob, and how to reformulate the Schwinger model into a qubit Hamiltonian. In section \ref{Sec_3_method} we discuss our method of obtaining results, including how we chose the initial variation parameter values, our choice of mixers, and CD term fixing. In section \ref{Sec_4_results_discussion} we show the results of the algorithms in finding the Schwinger model ground state for two different mixers and discuss them. Finally, we conclude on our results in section \ref{Sec_5_conclusion}. Throughout this paper, we use natural units, setting $\hbar = c = 1$.

\section{Background} \label{Sec_2_background}

\subsection{QAOA}

QAOA \cite{farhi2014} is a variational classical-quantum hybrid algorithm that is motivated by the Quantum Adiabatic Algorithm (QAA) \cite{farhi2000}. The adiabatic theorem \cite[pp.328-331]{Sakurai:2011} states that a quantum system will remain in its instantaneous eigenstate if the Hamiltonian of the system has a discrete, non-degenerate spectrum and is evolving slowly enough. How slow one needs to evolve the Hamiltonian is dependent on the energy gap between the eigenvalue of the eigenstate you want to remain in and the other eigenvalues of the Hamiltonian. QAA utilises this principle to find ground states of Hamiltonians. Consider a time-independent Hamiltonian $\h{H}_P$ (referred to as the problem Hamiltonian) whose ground we want to find, and another time-independent Hamiltonian $\h{H}_M$ (referred to as a mixer or mixer Hamiltonian) with a known ground state $|g\rangle$ that can easily be prepared. We can prepare the ground state of $\h{H}_P$ from the known ground of $\h{H}_M$ by starting a system in state $|g\rb$ and evolving under the Hamiltonian
\begin{equation}
    \h{H}_{QAA}(\lambda(t)) = (1-\lambda(t))\h{H}_M + \lambda(t)\h{H}_P,
\end{equation}
where $\lambda(t)\in[0,1]$ for $t\in[0,T]$ and $T$ is the total evolution time. If we control $\lambda(t)$ such that the adiabatic theorem applies then at time $t=T$ the system should be in the ground state of $\h{H}_P$. It is normally imposed that $\h{H}_P$ and $\h{H}_M$ do not commute to allow for the eigenbasis of $\h{H}_{QAA}(\lambda(t))$ to change during the evolution.\\

QAA requires a continuous evolution under $\h{H}_{QAA}$, represented by the operator
\begin{equation}
    \h{U}_{QAA}(t) = \mathcal{T}\left[e^{-i\int_{0}^{t} \h{H}_{QAA}(\lambda(\tau)) \ d\tau}\right].
\end{equation}
where $\mathcal{T}$ indicates that the exponential is time-ordered. The state of current hardware that can implement this kind of evolution (in the form of quantum annealers \cite{Vinci_2017,Rajak_2022}) is very limited, with $\h{H}_P$ often being restricted to the form of a classical Ising model Hamiltonian \cite{Amin_2011}. Instead, we look toward gate-based quantum computing which is more universal.\\ 

We can approximate the continuous evolution from QAA using quantum gates through Trotterization where we split $\h{U}_{QAA}$ into small time steps and assume $\h{H}_{QAA}$ is approximately constant over each step so that we can remove the time integral and time-ordering. If we set $\lambda(t)=\frac{t}{T}$ and $\Delta\tau=\frac{T}{p}$, where $p$ is a positive integer, we can approximate $\h{U}_{QAA}$ as \cite[pp. 9-10]{Blekos_2024}
\begin{equation}
\begin{split} \label{trot_QAA}
    \hat{U}_{QAA}(t) \approx& \prod_{k=0}^{p-1} \exp\left[-i\hat{H}_{QAA}(k\Delta\tau)\Delta\tau\right],\\ \approx& \prod_{k=0}^{p-1}\exp\left[ -i\left(1 - \frac{k\Delta\tau}{T} \right)\hat{H}_M \Delta\tau \right]\\&\times \exp\left[ -i\frac{k\Delta\tau}{T}\hat{H}_P\Delta\tau\right].
\end{split}
\end{equation}
where we have used the Lie-Trotter-Suzuki formula $e^{x(\hat{A}+\hat{B})} = e^{x\hat{A}}e^{x\hat{B}} + \mathcal{O}(x^2)$ \cite{Hatano_2005}. Splitting the evolution into small time steps makes it more practical to implement on a quantum computer using fundamental gate sets.\\

The starting point of QAOA is the construction of an ansatz state $|\boldsymbol{\gamma},\boldsymbol{\beta}\rb$ which is motivated by the Trotterize $\hat{U}_{QAA}$ in (\ref{trot_QAA}). We again start with a problem Hamiltonian $\h{H}_P$ and a mixer Hamiltonian $\h{H}_M$ with ground state $|g\rb$. We use these Hamiltonians to define the unitaries
\begin{equation}
    \h{U}_P(\gamma) = e^{-i\gamma\h{H}_P}, \quad
    \h{U}_M(\beta) = e^{-i\beta\h{H}_M},
\end{equation}
where $\gamma$ and $\beta$ are variational parameters. $|\boldsymbol{\gamma},\boldsymbol{\beta}\rb$ is then constructed by applying $\h{U}_P(\gamma)$ then $\h{U}_M(\beta)$ to the state $|g\rb$ and repeating this $p$ times to get
\begin{equation}
\begin{split}
    |\boldsymbol{\gamma},\boldsymbol{\beta}\rb &= \h{U}_M(\beta_p)\h{U}_P(\gamma_p)\cdots\h{U}_M(\beta_1)\h{U}_P(\gamma_1)|g\rb\\& = e^{-i\beta_p\h{H}_M}e^{-i\gamma_p\h{H}_P}\cdots e^{-i\beta_1\h{H}_M}e^{-i\gamma_1\h{H}_P}|g\rb
\end{split}
\end{equation}
where we now have the two sets of variational parameters $\boldsymbol{\gamma}=(\gamma_1,\gamma_2,\cdots\gamma_p)$ and $\boldsymbol{\beta}=(\beta_1,\beta_2,\cdots\beta_p)$. We note the similarity between the above and (\ref{trot_QAA}), with the unitaries in $|\boldsymbol{\gamma},\boldsymbol{\beta}\rb$ acting as a different parametrization of the Trotterized $\h{U}_{QAA}$. Each time the unitaries $\h{U}_P(\gamma)$ and then $\h{U}_M(\beta)$ are applied in the ansatz state, it is referred to as a QAOA layer and $p$ is the number of layers. Higher $p$ values correspond to a more accurate Trotterization of $\h{U}_{QAA}$ and should lead to QAOA performing better.\\

\begin{figure*}[!]
    \centering
\resizebox{1.8\columnwidth}{!}{
\begin{tabular}[b]{c}
{\hspace{-0.9cm}\large(a) QAOA} \hspace{6cm} {\large(b) DC-QAOA}\\
{\includegraphics[width=0.375\textwidth]{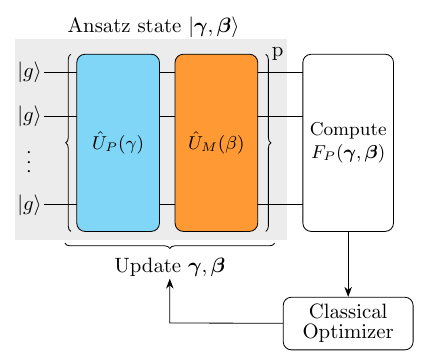}} \qquad{\includegraphics[width=0.48\textwidth]{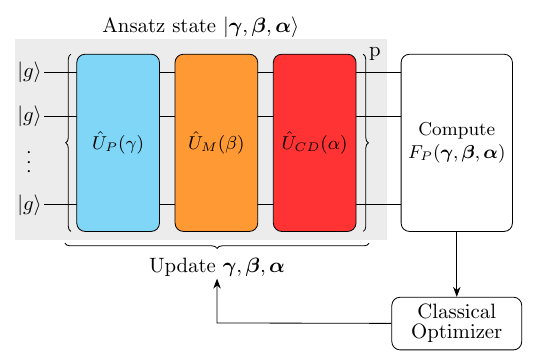}}
\end{tabular}}

\vspace{0.2cm}

\resizebox{2\columnwidth}{!}{
\begin{tabular}[b]{c}
{\hspace{-1.5cm}\large(c) CD-inspired} \hspace{3.75cm} {\large(d) CD-mixer} \hspace{4.5cm} {\large(e) CD-prob}\\
{\includegraphics[width=0.3\textwidth]{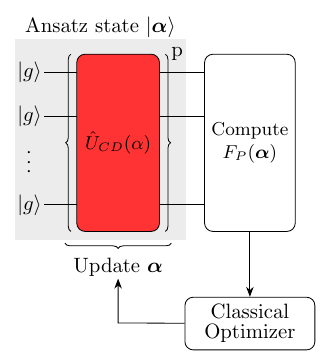}} {\includegraphics[width=0.39\textwidth]{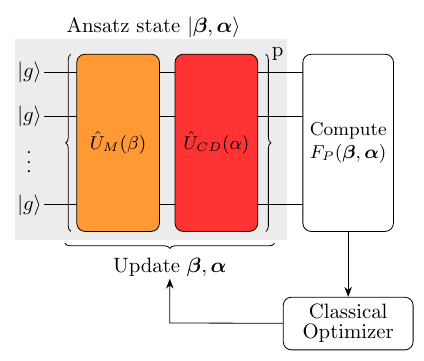}}
{\includegraphics[width=0.39\textwidth]{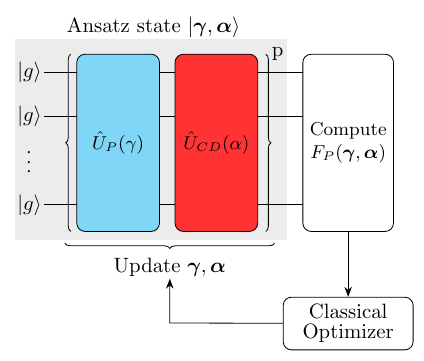}}
\end{tabular}}

\caption{The circuit diagrams for the algorithms QAOA in (a), DC-QAOA in (b), CD-inspired in (c), CD-mixer in (d) and CD-prob in (e). The expectation values $F_P(\boldsymbol{\gamma},\boldsymbol{\beta})$, $F_P(\boldsymbol{\gamma},\boldsymbol{\beta},\boldsymbol{\alpha})$, $F_P(\boldsymbol{\alpha})$, $F_P(\boldsymbol{\beta},\boldsymbol{\alpha})$ and $F_P(\boldsymbol{\gamma},\boldsymbol{\alpha})$ are defined in (\ref{QAOA_EV}), (\ref{DC-QAOA_EV}), (\ref{CD-inspired_EV}), (\ref{CD-mixer_EV}) and (\ref{CD-prob_EV}) respectively.}
\label{Circuit_diagrams}

\end{figure*}

The state $|\boldsymbol{\gamma},\boldsymbol{\beta}\rb$ is then used to calculate the expectation value
\begin{equation} \label{QAOA_EV}
    F_p(\boldsymbol{\gamma},\boldsymbol{\beta}) = \lb\boldsymbol{\gamma},\boldsymbol{\beta}|\h{H}_P|\boldsymbol{\gamma},\boldsymbol{\beta}\rb.
\end{equation}
This can be done on an actual quantum computer by estimating the expectation values of the Pauli strings that compose $\h{H}_P$ via repeated measurements in the appropriate bases, or it can be calculated exactly on a classical simulator. A classical optimizer is then used to continually update the variational parameters such that the optimal set of parameters $\boldsymbol{\gamma}^*,\boldsymbol{\beta}^*$ obey
\begin{equation}
    \boldsymbol{\gamma}^*,\boldsymbol{\beta}^* = \arg\min_{\boldsymbol{\gamma},\boldsymbol{\beta}} F_p(\boldsymbol{\gamma},\boldsymbol{\beta}).
\end{equation}
The more $\boldsymbol{\gamma},\boldsymbol{\beta}$ get optimized, the closer the energy eigenvalue of $|\boldsymbol{\gamma},\boldsymbol{\beta}\rb$ will be to the ground energy of $\h{H}_P$, and this should lead to $|\boldsymbol{\gamma},\boldsymbol{\beta}\rb$ becoming a more accurate approximate $\h{H}_P$ ground state. Optimizing the parameters in this way allows the evolution process created by the unitaries in $|\boldsymbol{\gamma},\boldsymbol{\beta}\rb$ to go beyond the standard adiabatic process and potentially find more optimal paths. A circuit diagram of QAOA is shown in Figure \ref{Circuit_diagrams}(a). \\

In its original proposal, QAOA was used to solve combinatorial optimization problems by encoding the problem in $\h{H}_P$ and having its ground state correspond to the optimal solution of the problem. In this context, it was assumed that $\h{H}_P$ was diagonal in the computational basis, with the bit strings that labelled the computational basis states corresponding to a classical solutions of the optimization problem. This made QAOA particularly powerful as $\h{H}_M$ could be chosen such that $|g\rb$ was a superposition of all computational basis states, and thus all possible solutions of the problem. Additionally, actual quantum computers are designed to measure in the computational basis making these problems natural to run on these devices. However, QAOA can still be applied to non-diagonal Hamiltonians but now the interpretation of the algorithm is different, with it being viewed more as a Variational Quantum Eigensolver \cite{Peruzzo_2014} with a certain ansatz.

\subsection{DC-QAOA}

DC-QAOA \cite{Chandarana_2022} is a proposed variation of QAOA that involves CD driving with the aim of reducing circuit depth by having improved results at a lower number of layers. Consider some system that is in an eigenstate $|\psi(t)\rb$ of the Hamiltonian $\h{H}_0(\lambda(t))$. We want to evolve the system under $\h{H}_0(\lambda(t))$ such that it remains in $|\psi(t)\rb$ throughout the evolution, which would normally require a slow evolution process. We can apply CD driving to the system by adding an extra term to $\h{H}_0$ so that we have
\begin{equation}
    \h{H}_{CD} = \h{H}_0 + \dot{\lambda}\h{A}_{\lambda}
\end{equation}
where $\h{A}_\lambda$ is an object known as an adiabatic gauge potential \cite{Kolodrubetz_2017}. The term $\dot{\lambda}\h{A}_{\lambda}$ acts to exactly cancel transitions to other eigenstates during the evolution, allowing for the system to remain in  $|\psi(t)\rb$ with no requirements for slow evolution.\\

In practise implementing the exact CD term is usually very difficult due to $\h{A}_\lambda$ involving non-local many-body terms and requiring full knowledge of the spectrum of $\h{H}_0$. Instead an approximate $\h{A}_\lambda$ term can be found which does not require spectral knowledge of $\h{H}_0$, allowing for approximate CD driving term which will help to suppress transitions to other eigenstates. A common way to approximate $\h{A}_\lambda$ is the nested commutator approach \cite{Claeys_2019} where $\h{A}_\lambda$ is given as the order expansion
\begin{equation} \label{NC_AGP}
    \hat{A}^{(\ell)}_\lambda = i \sum_{k=1}^\ell \alpha_k(t) \underbrace{[\hat{H}_0,[\hat{H}_0,...[\hat{H}_0,}_{2k-1} \partial_{\lambda}\hat{H}_0]]]
\end{equation}
where $\ell$ is the order of the expansion and $\alpha_k(t)$ are a set of coefficients. The above tends to the exact $\h{A}_\lambda$ expression as $\ell\rightarrow \infty$.\\ 

In DC-QAOA, $\hat{A}^{(\ell)}_\lambda$ is further simplified. The nested commutators result in operator terms which are summed over the qubits in certain ways. These operator terms make up an operator pool $A$, and instead of implementing all the operators in the pool which would be costly, one is heuristically chosen from it and then summed over the qubits. We will refer to this summed term as $\h{A}_{DC}$. In DC-QAOA, $\h{A}_{DC}$ is used to define the unitary \begin{equation}
    \h{U}_{CD}(\alpha) = e^{-i\alpha\h{A}_{DC}},
\end{equation}
where $\alpha$ is a variational parameter. This unitary contains the CD effects.\\

To physical motivate the DC-QAOA ansatz state, we set $\h{H}_0(\lambda(t))=\h{H}_{QAA}(\lambda(t))$ and make the replacement $\h{A}_\lambda\rightarrow \h{A}_{DC}$ so that we have the Hamiltonian 
\begin{equation} \label{CD_QAA_Ham}
    \h{H}_{CD,QAA} = (1-\lambda(t))\h{H}_M + \lambda(t)\h{H}_P + \dot{\lambda}\h{A}_{DC},
\end{equation}
then the evolution operator for evolving under $\h{H}_{CD,QAA}$ is given by
\begin{equation}
\begin{split}
    \h{U}_{CD,QAA}(t) &= \mathcal{T}\left[e^{-i\int_{0}^{T} \h{H}_{CD,QAA}(\lambda(t)) \ dt}\right].
\end{split}
\end{equation}
Just as the QAOA ansatz state was motivated by the Trotterized $\h{U}_{QAA}(t)$, the Trotterized $\h{U}_{CD,QAA}(t)$ leads to the DC-QAOA ansatz state of
\begin{equation}
    \begin{split}    |\boldsymbol{\gamma},\boldsymbol{\beta},\boldsymbol{\alpha}\rb = &\ \h{U}_{CD}(\alpha_p)\h{U}_M(\beta_p)\h{U}_P(\gamma_p)\cdots \\ &\ \h{U}_{CD}(\alpha_1) \h{U}_M(\beta_1) \h{U}_P(\gamma_1)|g\rb\\ = &\ e^{-i\alpha_p\h{A}_{DC}}e^{-i\beta_p\h{H}_M}e^{-i\gamma_p\h{H}_P}\cdots \\ &\ e^{-i\alpha_1\h{A}_{DC}}e^{-i\beta_1\h{H}_M}e^{-i\gamma_1\h{H}_P}|g\rb
\end{split}
\end{equation}
where we now have an extra set of parameters $\boldsymbol{\alpha}=(\alpha_1,\alpha_2,\cdots\alpha_p)$. The rest of the algorithm is modified accordingly. We now calculate the expectation value
\begin{equation} \label{DC-QAOA_EV}
    F_p(\boldsymbol{\gamma},\boldsymbol{\beta}, \boldsymbol{\alpha}) = \lb\boldsymbol{\gamma},\boldsymbol{\beta}, \boldsymbol{\alpha}|\h{H}_P|\boldsymbol{\gamma},\boldsymbol{\beta}, \boldsymbol{\alpha}\rb,
\end{equation}
and the classical optimizer optimizes the parameters such that the optimal set of parameters $\boldsymbol{\gamma}^*,\boldsymbol{\beta}^*,\boldsymbol{\alpha}^*$ obey
\begin{equation}
    \boldsymbol{\gamma}^*,\boldsymbol{\beta}^*,\boldsymbol{\alpha}^* = \arg\min_{\boldsymbol{\gamma},\boldsymbol{\beta},\boldsymbol{\alpha}} F_p(\boldsymbol{\gamma},\boldsymbol{\beta},\boldsymbol{\alpha}).
\end{equation}
Even though an additional set of parameters are added, which can make optimization difficult, the aim is for the CD effects to allow the optimizer to find more efficient paths towards the $\h{H}_P$ ground state and so require a reduced number of layers in the ansatz state compared to QAOA, lowering the number of variational parameters and circuit depth. A circuit diagram of DC-QAOA is shown in Figure \ref{Circuit_diagrams}(b).

\subsection{DC-QAOA variants}

As previously mentioned, by modifying the QAOA ansatz to include CD effects we have increased the number of variational parameters per layer and the depth of our circuit. This can make the optimization more difficult and the algorithm harder to implement on actual quantum hardware. This has led to further modifications of the DC-QAOA that involve removing one or both of the non-CD unitaries in each layer of the ansatz state. The motivation behind this is instead of implementing the full adiabatic evolution process, we can remove parts of it and rely on the classical optimizer to still lead us towards the ground state. If we retain the CD parts of the ansatz state we can still get the effects of the fast evolution that CD driving provides. The expectation value $F_p(\boldsymbol{\gamma},\boldsymbol{\beta}, \boldsymbol{\alpha})$ and optimization routines are modified accordingly by replacing $|\boldsymbol{\gamma},\boldsymbol{\beta},\boldsymbol{\alpha}\rb$ with the modified ansatz and reducing the number of parameters to optimize over.\\

The following DC-QAOA variant is called CD-inspired \cite{Chandarana_2023} and its ansatz state is
\begin{equation}
    \begin{split}    |\boldsymbol{\alpha}\rb &= \h{U}_{CD}(\alpha_p)\cdots \h{U}_{CD}(\alpha_1) |g\rb\\& = e^{-i\alpha_p\h{A}_{DC}}\cdots e^{-i\alpha_1\h{A}_{DC}}|g\rb,
\end{split}
\end{equation}
where only the counterdiabatic parts of the ansatz state are retained. The physical motivation behind this ansatz state is the case where $|\dot{\lambda}(t)|>>|\lambda(t)|$ in (\ref{CD_QAA_Ham}) so that the $\h{A}_\lambda$ term dominates and we can ignore the other terms, allowing for a simpler evolution process. The CD-inspired expectation value is
\begin{equation} \label{CD-inspired_EV}
    F_p(\boldsymbol{\alpha}) = \lb \boldsymbol{\alpha}|\h{H}_P|\boldsymbol{\alpha}\rb,
\end{equation}
and the classical optimizer optimizes the parameters such that the optimal set of parameters $\boldsymbol{\alpha}^*$ obey
    \begin{equation}  \boldsymbol{\alpha}^* = \arg\min_{\boldsymbol{\alpha}} F_p(\boldsymbol{\alpha}).
\end{equation}\\

We also have CD-mixer \cite{Xu_2025} where the ansatz state includes the CD and mixer unitaries
\begin{equation}
    \begin{split}    |\boldsymbol{\beta},\boldsymbol{\alpha}\rb =& \h{U}_{CD}(\alpha_p)\h{U}_M(\beta_p)\cdots \h{U}_{CD}(\alpha_1) \h{U}_M(\beta_1)|g\rb\\ =& e^{-i\alpha_p\h{A}_{DC}}e^{-i\beta_p\h{H}_M}\cdots \\& e^{-i\alpha_1\h{A}_{DC}}e^{-i\beta_1\h{H}_M}|g\rb.
\end{split}
\end{equation}
Including the mixer in the ansatz state allows it to be more expressive, and may start the optimization process closer to the global minimum due to the mixer being involved in the CD evolution. This is without adding too many gates to the circuit or too much complexity to the optimization process due to mixers generally being simple. The CD-mixer expectation value is
\begin{equation} \label{CD-mixer_EV}
    F_p(\boldsymbol{\beta}, \boldsymbol{\alpha}) = \lb\boldsymbol{\beta}, \boldsymbol{\alpha}|\h{H}_P|\boldsymbol{\beta}, \boldsymbol{\alpha}\rb,
\end{equation}
and the classical optimizer optimizes the parameters such that the optimal set of parameters $\boldsymbol{\beta}^*,\boldsymbol{\alpha}^*$ obey
\begin{equation}
    \boldsymbol{\beta}^*,\boldsymbol{\alpha}^* = \arg\min_{\boldsymbol{\beta},\boldsymbol{\alpha}} F_p(\boldsymbol{\beta},\boldsymbol{\alpha}).
\end{equation}\\

Finally, for completeness, we have included a new algorithm we call CD-prob, which has an ansatz where we only remove the mixer unitaries
\begin{equation}
    \begin{split}    
    |\boldsymbol{\gamma},\boldsymbol{\alpha}\rb =& \h{U}_{CD}(\alpha_p)\h{U}_P(\gamma_p)\cdots \h{U}_{CD}(\alpha_1) \h{U}_P(\gamma_1)|g\rb\\ =& e^{-i\alpha_p\h{A}_{DC}}e^{-i\gamma_p\h{H}_P}\cdots \\&e^{-i\alpha_1\h{A}_{DC}}e^{-i\gamma_1\h{H}_P}|g\rb.
\end{split}
\end{equation}
In this case, $\h{A}_{DC}$ is essentially acting as the mixer but it is now more physically motivated. In the algorithm ADAPT-QAOA \cite{Zhu_2022}, where the mixer is chosen from an operator pool based on how well it performs, often the mixers chosen for the lowest layers are the same as the operators with the largest coefficient in the pool $A$ generated by (\ref{NC_AGP}). The CD-prob expectation value is
\begin{equation} \label{CD-prob_EV}
    F_p(\boldsymbol{\gamma}, \boldsymbol{\alpha}) = \lb\boldsymbol{\gamma}, \boldsymbol{\alpha}|\h{H}_P|\boldsymbol{\gamma}, \boldsymbol{\alpha}\rb,
\end{equation}
and the classical optimizer optimizes the parameters such that the optimal set of parameters $\boldsymbol{\gamma}^*,\boldsymbol{\alpha}^*$ obey
\begin{equation}
    \boldsymbol{\gamma}^*,\boldsymbol{\alpha}^* = \arg\min_{\boldsymbol{\gamma},\boldsymbol{\alpha}} F_p(\boldsymbol{\gamma},\boldsymbol{\alpha}).
\end{equation}\\

We refer to these three algorithms, along with DC-QAOA, as the CD algorithms, to distinguish them from QAOA. Circuit diagrams of CD-inspired, CD-mixer and CD-prob are shown in Figures \ref{Circuit_diagrams}(c), \ref{Circuit_diagrams}(d) and \ref{Circuit_diagrams}(e) respectively.

\subsection{The Schwinger model}

The Schwinger model is a 1+1 dimensional $U(1)$ gauge theory coupled to a Dirac fermion \cite{Schwinger_1962_1,Schwinger_1962_2}. The Lagrangian for this model is
\begin{equation}
\begin{split}
    \mathcal{L}_{Schwinger} =& \frac{1}{4}F_{\mu\nu}F^{\mu\nu} + \frac{g\theta}{4\pi}\epsilon_{\mu\nu}F^{\mu\nu}\\& + i\bar{\psi}\gamma^\mu(\partial_\mu + igA_\mu)\psi - m\bar{\psi}\psi,
\end{split}
\end{equation}
where $F_{\mu\nu}=\partial_\mu A_\nu-\partial_\nu A_\mu$, $g$ is the gauge coupling, $\theta$ is the topological angle \cite{izubuchi2008}, $\epsilon_{\mu\nu}$ is the Levi-Civita symbol, $\psi$ is the Dirac spinor, $\bar{\psi}=\psi^\dagger\gamma^0$ is the Dirac adjoint, $A_\mu$ is the electromagnetic four-potential, and $m$ is the fermion mass.\\ 

We want to reformulate this model as a Hamiltonian that acts on qubits so we can run it on a quantum computer (full details on how to do this in \cite{chakraborty2022}). First we regularize the infinite Hilbert space by placing the theory on a spatial lattice using staggered fermions to discretize the spatial dimension (while keeping time continuous) and imposing the Gauss law which gives the Hamiltonian
\begin{equation} \label{Schwinger_gauge_Ham}
\begin{split}
    \h{H}_{Schwinger} =& -i\sum_{n=1}^{N-1} \left( w - (-1)^n\frac{m}{2}\sin\theta \right) \\&\times\left[\h{\chi}_n^\dagger e^{i\h{\phi}_n}\h{\chi}_{n+1} - \h{\chi}_{n+1}^\dagger e^{-i\h{\phi}_n}\h{\chi}_n\right]\\& + m\cos\theta\sum_{n=1}^N(-1)^n\h{\chi}_n^\dagger\h{\chi}_n + J\sum_{n=1}^{N-1}\h{L}_n^2,
\end{split}
\end{equation}
where $\h{\chi}_n$ is the staggered fermion, $N$ is the number of lattice sites, $a$ is the lattice spacing, $w=\frac{1}{2a}$, $J=\frac{g^2a}{2}$, $\h{\phi}_n\leftrightarrow -ag\h{A}^1(x)$ is a rescaled gauge operator which lives on site $n$, and $\h{L}_n\leftrightarrow -\frac{\h{\Pi}(x)}{g}$ is a rescaled gauge operator, where $\h{\Pi}(x)$ is the $\h{A}^1(x)$ conjugate momentum, which lives on the link between sites $n$ and $n+1$.\\

Now we need to convert the operators in (\ref{Schwinger_gauge_Ham}) to easily implementable quantum gates. A common set of gates are the Pauli gates with the identity gate $\{\X, \Y, \Z, \h{I}\}^{\otimes n}$ where
\begin{equation}
\begin{split}
    &\hat{X} = \begin{pmatrix} 0&1\\1&0\end{pmatrix}, \quad
    \hat{Y} = \begin{pmatrix} 0&-i\\i&0\end{pmatrix}, \\
    &\hat{Z} = \begin{pmatrix} 1&0\\0&-1\end{pmatrix},
    \quad
    \h{I} = \begin{pmatrix} 1&0\\0&1\end{pmatrix},
\end{split}
\end{equation}
which can be used to express any operator acting on a qubit Hilbert space. We can map the $\h{\chi}_n$ operators to Pauli gates using the Jordan-Wigner transformation
\begin{equation}
    \h{\chi}_n = \left(\prod_{l<n}-i\h{Z}_l\right)\frac{\h{X}_n-i\h{Y}_n}{2}.
\end{equation}
We can map the $\h{L}_n$ operators to Pauli gates by imposing an open boundary condition which restricts $\h{L}_n$ to a constant at the boundary and solving the Gauss law to get
\begin{equation}
    \h{L}_n = \frac{1}{2}\sum_{l=1}^n\left(\h{Z}_l+(-1)^l\right).
\end{equation}
Finally, we can eliminate the $\h{\phi}_n$ operators using the redefinition $\h{\chi}_n\rightarrow\Pi_{\ell<n}\left[e^{-i\h{\phi}_\ell}\right]\h{\chi}_n$.\\

This gives the lattice Schwinger model in turns of Pauli gates as
\begin{equation}
    \hat{H}_{Schwinger} = \hat{H}_{ZZ} + \hat{H}_\pm + \hat{H}_Z,
\end{equation}
where
\begin{gather}
    \hat{H}_{ZZ} = \frac{J}{2}\sum_{n=2}^{N-1}\sum_{1\leq k<l\leq n}\hat{Z}_k\hat{Z}_l,\\
\begin{split}
    \hat{H}_\pm =&\ \frac{1}{2}\sum_{n=1}^{N-1}\left(w-(-1)^n\frac{m\sin\theta}{2}\right)\\ &\ \times\left[\hat{X}_n\hat{X}_{n+1} + \hat{Y}_n\hat{Y}_{n+1}\right],
\end{split}\\
\begin{split}
    \hat{H}_Z =&\ \frac{m\cos\theta}{2}\sum_{n=1}^N(-1)^n\hat{Z}_n \\&\ - \frac{J}{2}\sum_{n=1}^{N-1}(n\text{ mod } 2)\sum_{l=1}^n \Z_l.
\end{split}
\end{gather}
The Schwinger model formulated in this way can be ran on a quantum computer, and quantum algorithms can be used to find its ground state. When we run this Hamiltonian on a quantum computer, the lattice sites will correspond to individual qubits with the indices on the operators indicating which qubit they are acting on.\\

Once the ground state of $\h{H}_{Schwinger}$ is found, it can be used to calculate some observable e.g. a vacuum expectation value or correlation function. One can then return to the continuous theory by taking the continuum limit $a\rightarrow 0$ for these observables. The Schwinger model also has a finite mass gap \cite{Ba_uls_2013}, except at a critical point of $\theta=\pi$ and $m$ equal to some critical mass, where a phase transition occurs and the gap closes \cite{Cruz_2025}. However, away from this critical point, the use of the adiabatic theorem is valid and the counterdiabatic operators remain well defined in the continuum limit as the energy spectrum remains gapped. 

\section{Method} \label{Sec_3_method}

The algorithms QAOA, DC-QAOA, CD-inspired, CD-mixer and CD-prob were used to prepare the ground state of $\h{H}_{Schwinger}$ by running them with $\h{H}_P=\h{H}_{Schwinger}$. The algorithms were coded using Qiskit 2.2.3 \cite{javadiabhari2024} and simulated on Qiskit Aer 0.17.2 using AerSimulator which measured expectation values exactly. The code used to obtain the results in this paper is publicly available on the GitHub repository \href{https://github.com/EthanLaval/Counterdiabatic-quantum-optimization-for-efficient-state-preparation-in-the-Schwinger-model}{EthanLaval/Counterdiabatic-quantum-optimization-for-efficient-state-preparation-in-the-Schwinger-model}.\\

The classical optimizer used in the algorithms was GradientDescent from the Qiskit Algorithms library, with maxiter set to 300 and the rest of the settings set to default. After the parameters were optimized, the success of the algorithms was measured by taking the final expectation value $F_p^*$ of the algorithm after 300 iterations of the optimizer, and calculating the following approximation ratio
\begin{equation}
    \mathcal{R} = \frac{F_p^*}{E_{Schwinger}}
\end{equation}
where $E_{Schwinger}$ is the ground energy of the $\h{H}_{Schwinger}$ found through exact diagonalization. Another measurement that was looked at was the state fidelity between the final ansatz state $|A^*\rb$ outputted by the algorithms and the $\h{H}_{Schwinger}$ ground state $|g_S\rb$ found through exact diagonalization, which is given by
\begin{equation}
    \mathcal{F} = |\lb A^*|g_{S}\rb|^2.
\end{equation}
 We ran each of the algorithms multiple times so the average of the approximation ratios and state fidelities found in each run was taken.\\

By finding a state that has on average an energy eigenvalue close to the ground energy of $\h{H}_{Schwinger}$, we can argue that this state is an approximate ground state of $\h{H}_{Schwinger}$ and further validate this with the state fidelity. The Schwinger model parameters were chosen to be $m = 0.35$ so that we are working in an intermediate mass regime, $a = 5/6$ as a convenient lattice spacing, $g = 1$ to set the energy scale, and $\theta = 0$ to simplify the physics and focus on the ground state preparation rather than the $\theta$ dependence of the model. The algorithms were ran for 1 to 5 layers and 4, 6, 8 and 10 qubits.

\subsection{Parameter fixing strategy} \label{subsec: Parameter fixing strategy}

The performance of QAOA can be very depended on the initial values of the variational parameters. Often QAOA is ran multiple times with different randomly chosen initial parameters and an average is taken of the results. The problem with this approach is that as you increase the number of qubits in the algorithm the optimizer tends to get stuck in local minima causing the algorithm to scale poorly.\\

To help mitigate this problem we employed the parameter fixing strategy proposed in \cite{Lee_2021}. The strategy involves building up the algorithms layer by layer. Starting with the first layer, the algorithms are ran 20 times with different random initial values for the parameters. The set of optimized parameters at the end of the run that reached the highest approximation ratio are then passed into the ansatz state with two layers as the initial values for the parameters in the first layer. The initial parameters in the second layer are then randomized, and the algorithms are ran 20 times and then the optimized parameters at the end of the run with the highest approximation ratio are passed into the ansatz state with three layers, and this process repeats. Note that this allows the optimized parameters found in the lower layer ansatz states to be re-optimized. This strategy often leads to higher average approximation ratios.\\

When we randomize the initial variational parameters values in our algorithms, we set them to random values between 0 and $2\pi$. The same randomized parameters were passed into all the algorithms. E.g. with 1 layer we had a randomized initial $\alpha$, $\beta$, and $\gamma$ values. The randomized $\alpha$ is passed into the $\h{U}_{CD}$ in the CD algorithms, the randomized $\beta$ is passed in the $\h{U}_M$ in QAOA, DC-QAOA and CD-mixer, and the randomized  $\gamma$ was passed in the $\h{U}_P$ in QAOA, DC-QAOA and CD-prob.

\subsection{Mixers} \label{Sec_3B_mixers}

The algorithms require a choice for the mixer Hamiltonian. In this paper we looked at two different mixers. One of the choices was the X mixer
\begin{equation}
    \hat{H}_{M,X} = \sum_{n=1}^{N} \hat{X}_n,
\end{equation}
which is the standard mixer used in QAOA. The X mixers is designed for Hamiltonians diagonal in the computational basis so it is useful to look at other mixers which may be better suited for our problem Hamiltonian. With this in mind, the other mixer we looked at is the XY mixer \cite{Hadfield_2019}
\begin{equation}
    \quad \hat{H}_{M,XY} = \frac{1}{2}\sum_{n=1}^{N} \hat{X}_n\hat{X}_{n+1} + \hat{Y}_n\hat{Y}_{n+1}.
\end{equation}
This mixer is more appropriate due to its similarity to the $\h{H}_{\pm}$ term in $\h{H}_{Schwinger}$, which makes it more likely that the adiabatic path between them is more natural due to them having similar physical interactions. This mixer was used in reference \cite{Tomlinson_2025} for the Schwinger model with good results. The downside to using this mixer instead of the X mixer is that it more complicated, leading to a larger gates count and depth in the circuit. Also, its ground state is much harder to prepare compared to the X mixer ground state, though this problem could be mitigated by using a similar, but simpler, state instead.\\

\renewcommand{\arraystretch}{2}

\begin{figure*}[!]
\centering

\begin{tabular}[b]{c}
\hspace{0.5cm}{\large(a) 4 qubits} \hspace{7cm} {\large(b) 6 qubits}\\
{\includegraphics[width=0.48\textwidth]{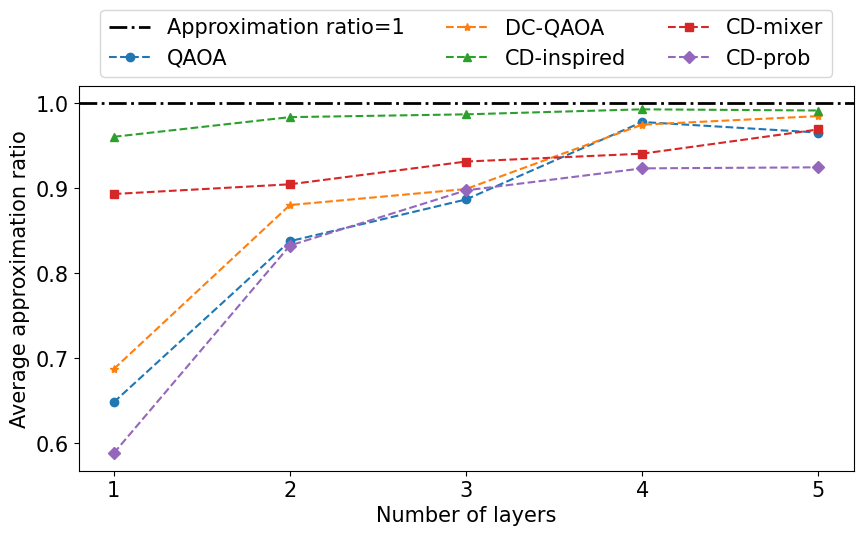}} \qquad{\includegraphics[width=0.48\textwidth]{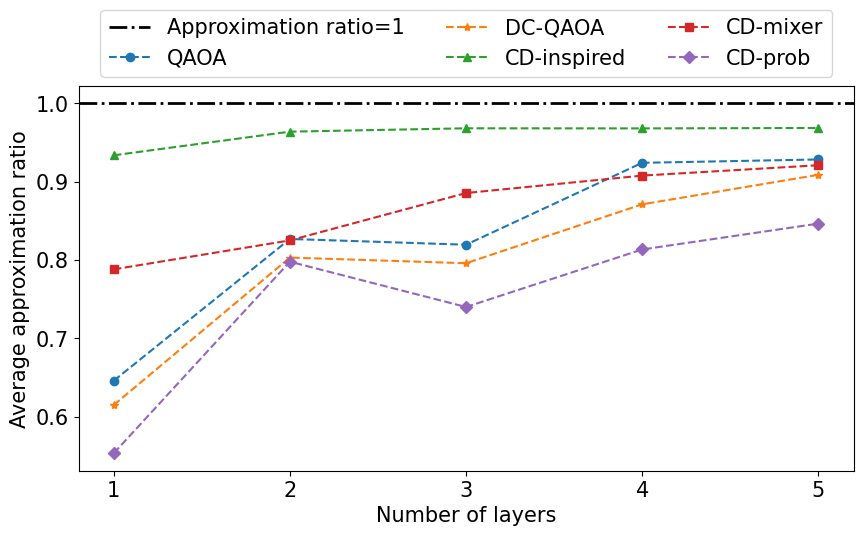}}
\end{tabular}

\begin{tabular}[b]{c}
\hspace{0.5cm}{\large(c) 8 qubits} \hspace{7cm} {\large(d) 10 qubits}\\
{\includegraphics[width=0.48\textwidth]{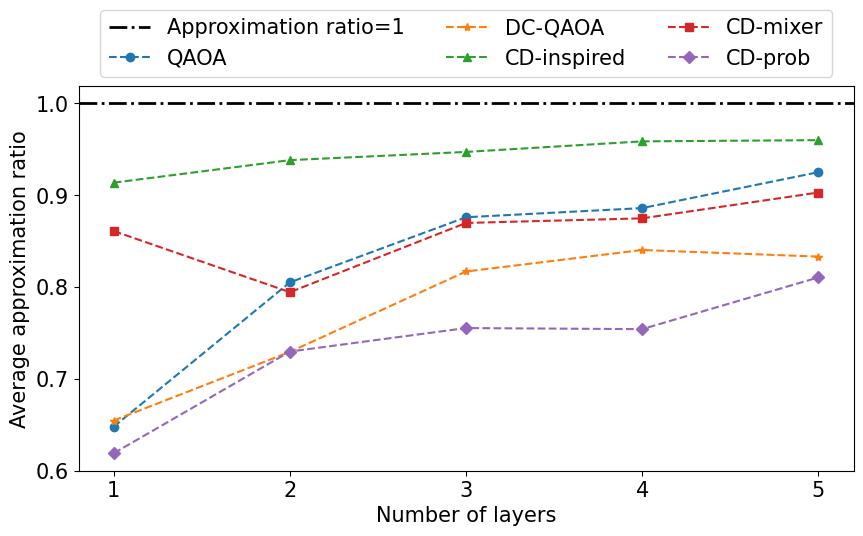}} \qquad{\includegraphics[width=0.48\textwidth]{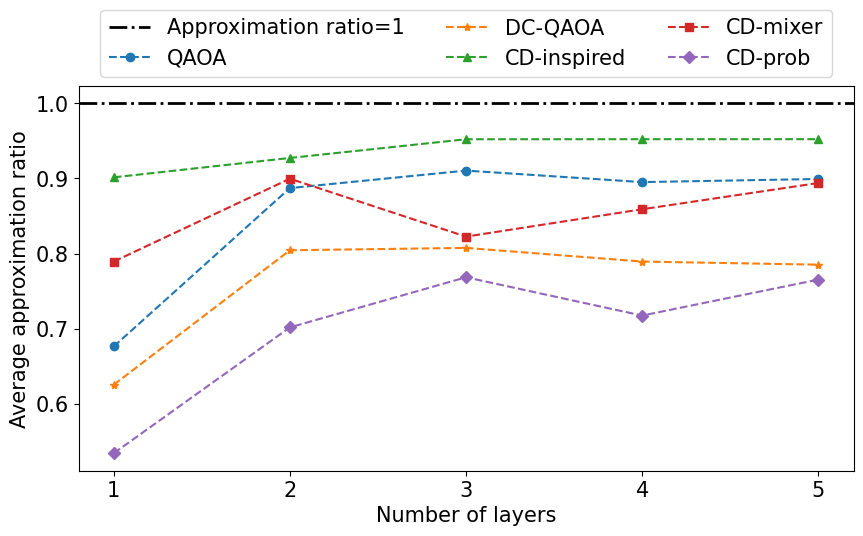}}
\end{tabular}

\caption{The average approximation ratios of the algorithms with an XY mixer for the Schwinger model against the number of layers for 4 qubits in (a), 6 qubits in (b), 8 qubits in (c), and 10 qubits in (d). The model parameters are $m = 0.35$, $w = 0.6$, $J = 5/12$ and $\theta = 0$.}
\label{XY_mixer}

\end{figure*}

\begin{figure*}[!]
\centering

\begin{tabular}[b]{c}
\hspace{0.5cm}{\large(a) 4 qubits} \hspace{7cm} {\large(b) 6 qubits}\\
{\includegraphics[width=0.48\textwidth]{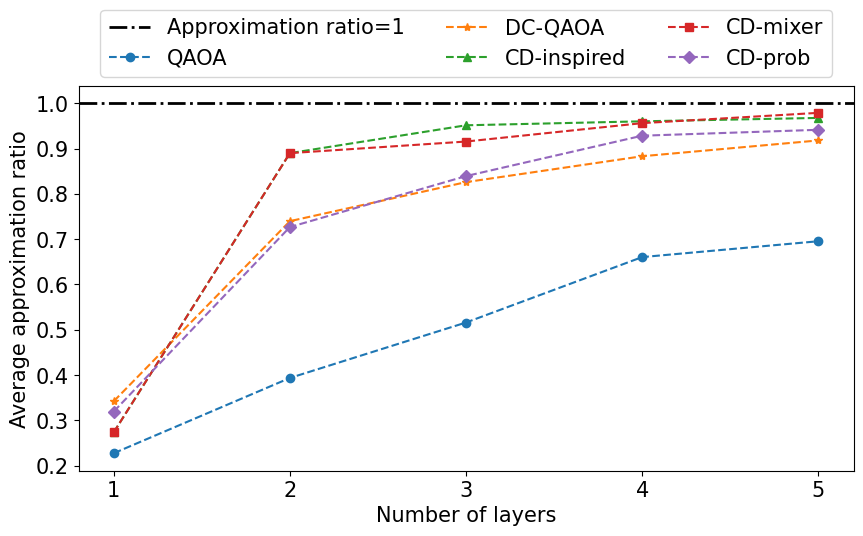}} \qquad{\includegraphics[width=0.48\textwidth]{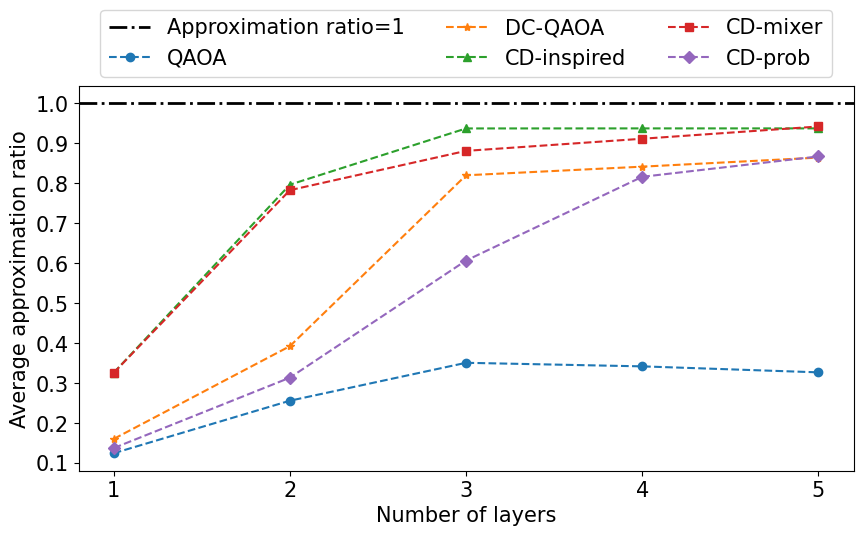}}
\end{tabular}

\begin{tabular}[b]{c}
\hspace{0.5cm}{\large(c) 8 qubits} \hspace{7cm} {\large(d) 10 qubits}\\
{\includegraphics[width=0.48\textwidth]{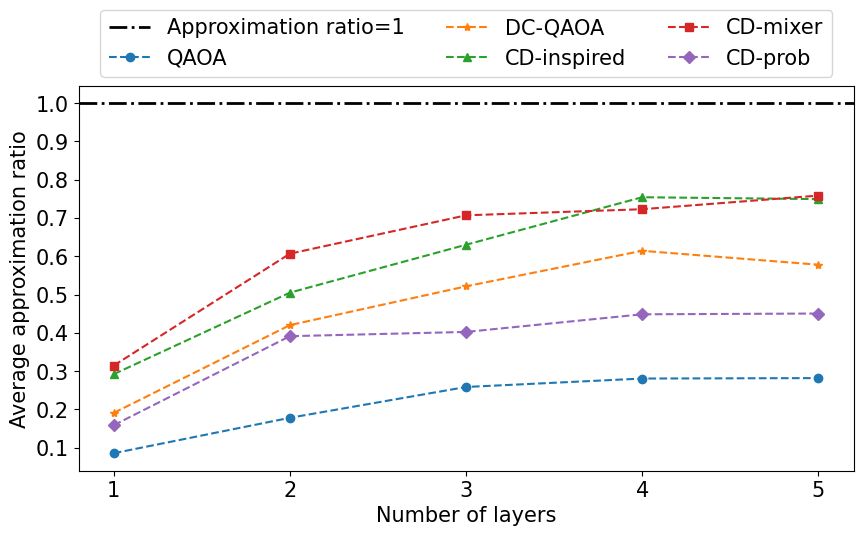}} \qquad{\includegraphics[width=0.48\textwidth]{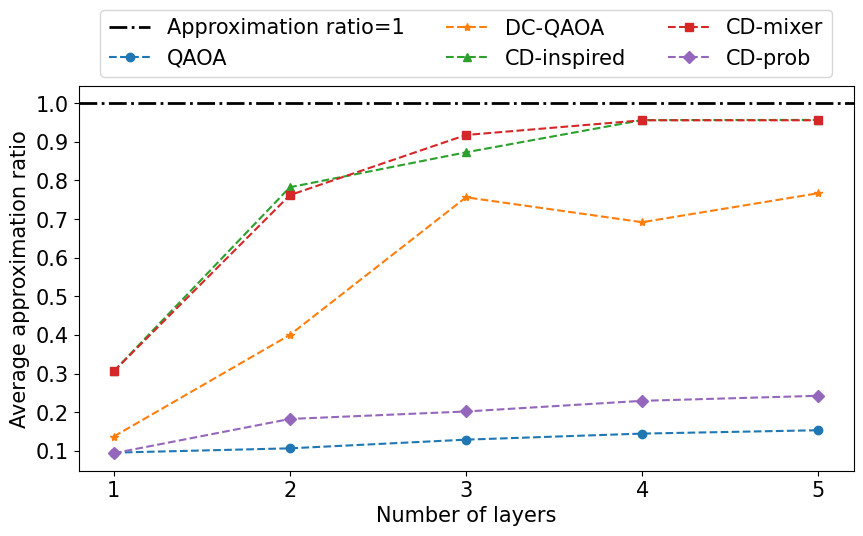}}
\end{tabular}

\caption{The average approximation ratios of the algorithms with an X mixer for the Schwinger model against the number of layers for 4 qubits in (a), 6 qubits in (b), 8 qubits in (c), and 10 qubits in (d). The model parameters are $m = 0.35$, $w = 0.6$, $J = 5/12$ and $\theta = 0$.}

\label{X_mixer}

\end{figure*}

\subsection{CD term fixing} \label{subsec: CD term fixing}

In the original DC-QAOA proposal, the $\h{A}_{DC}$ term is heuristically chosen from the operator pool $A$ generated by the nested commutators in (\ref{NC_AGP}). This term then fixes the unitary $\h{U}_{CD}$ for all the layers of the ansatz, just as $\h{H}_P$ and $\h{H}_M$ fixes $\h{U}_P$ and $\h{U}_M$ respectively. Similar to how the mixer is treated in ADAPT-QAOA \cite{Zhu_2022}, we allowed for the $\h{A}_{DC}$ term to be different in each layer of the ansatz states for our CD algorithms.\\

To chose which $\h{A}_{DC}$ term we have in each layer, we started with the first layer and we ran the algorithms with every possible operator in $A$ 20 times each. We then saw which $\h{A}_{DC}$ term resulted in the highest average approximation ratio and fix that as the $\h{A}_{DC}$ term for that layer. At the same time, we are using the parameter fixing strategy, so the initial parameters for that layer are the optimized parameters from the run with the fixed $\h{A}_{DC}$ term that had the highest approximation ratio. We then move onto the next layer and repeat this process with the $\h{A}_{DC}$ term fixed in the first layer and similarly for higher layers. By doing this we hope to localise the counterdiabatic effects and make full use of the operator pool in order to get higher approximation ratios. Though, this does come with the disadvantage that you need to run the CD algorithms a higher number of times compared to QAOA.\\

Which operators appears in the CD operator pools for our choice of $\h{A}_{DC}$ is determined by the mixer Hamiltonian. To calculate the CD operator pools, we set $\h{H}_0(\lambda(t))=\h{H}_{QAA}(\lambda(t))$ with $\h{H}_P=\h{H}_{Schwinger}$ and $\h{H}_M$ set to $\h{H}_{M,X}$ or $\h{H}_{M,XY}$ and used (\ref{NC_AGP}). If we go up to $\ell=2$ in (\ref{NC_AGP}) and only keep up to two-body operators, then setting $\h{H}_M = \hat{H}_{M,X}$ leads to the operator pool
\begin{equation}
    A_{X} = \{\hat{Y}, \hat{X}\hat{Y}, \hat{Y}\hat{X}, \hat{Y}\hat{Z}, \hat{Z}\hat{Y}\}.
\end{equation}
If we go up to $\ell=2$ and only keep up to three-body operators, then setting $\h{H}_M = \hat{H}_{M,XY}$ leads to the operator pool
\begin{equation}
\begin{split}
    A_{XY} =&\ \{\hat{X}\hat{Y}, \hat{Y}\hat{X}, \hat{X}\hat{Y}\hat{Z}, \hat{X}\hat{Z}\hat{Y}, \\&\ \hat{Y}\hat{X}\hat{Z}, \hat{Y}\hat{Z}\hat{X}, \hat{Z}\hat{X}\hat{Y}, \hat{Z}\hat{Y}\hat{X}  \}.
\end{split}
\end{equation}
We go up to three-body operators in this case just to have more operators in the pool. For simplicity, when we chose an operator term from $A$, we sum it over all the qubits while retaining the ordering of the operators to get $\h{A}_{DC}$. E.g. if we chose the $\X\Y$ operator term from $A_{XY}$ then $\h{A}_{DC}$ will be
\begin{equation}
    \h{A}_{DC} = \sum_{1\leq n<m\leq N} \X_n\Y_m.
\end{equation}\\

We conclude this section by referring the interested reader to \cite{cipolla2026} for an alternative approach to choose the CD term layer-adaptively.

\section{Results and discussion} \label{Sec_4_results_discussion}

Figure \ref{XY_mixer} shows the average approximation ratios from running the 5 different algorithms with an XY mixer for the Schwinger model using 1 to 5 layers and 4, 6, 8 and 10 qubits. We also present, in Figure \ref{X_mixer}, the results obtained using an X mixer. The average approximation ratios reported for the Counterdiabatic (CD) algorithms are the average approximation ratios obtained for the fixed   
CD term from the pool for that layer (calculated from 20 runs). And indeed, the results for the runs obtained by other CD terms were used solely for the CD term fixing procedure and were not included in the average. This gives both QAOA and the CD algorithms an equal number of runs (20) when calculating their average approximation ratios. For the sake of completeness, we also include in Appendix \ref{append: best data} the plots corresponding to the best approximation ratio found during the runs with the XY mixer.\\

\renewcommand{\arraystretch}{2}

\begin{figure*}[!]
\centering

\begin{tabular}[b]{c}
\hspace{0.5cm}{\large(a) 4 qubits} \hspace{7cm} {\large(b) 6 qubits}\\
{\includegraphics[width=0.48\textwidth]{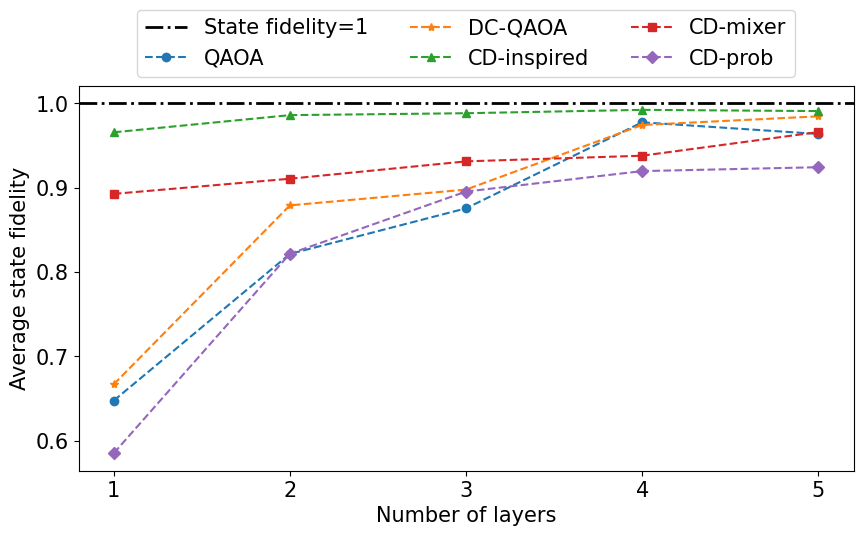}} \qquad{\includegraphics[width=0.48\textwidth]{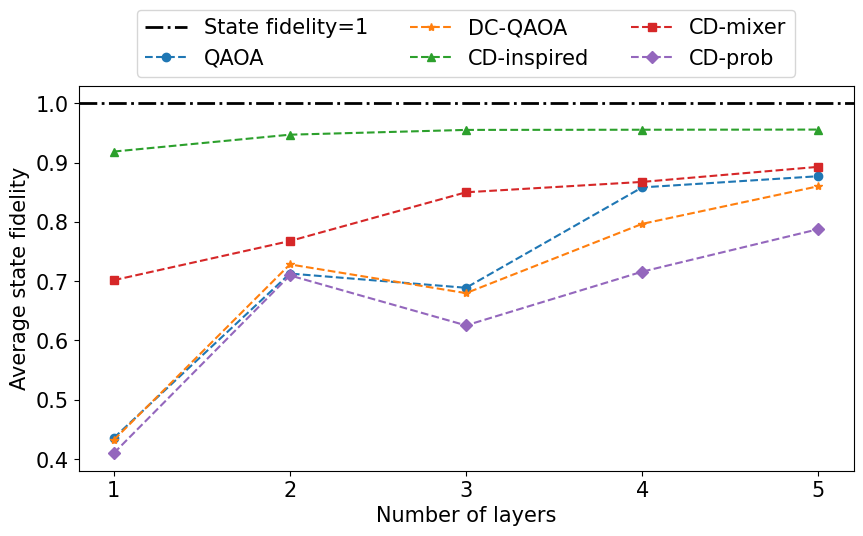}}
\end{tabular}

\begin{tabular}[b]{c}
\hspace{0.5cm}{\large(c) 8 qubits} \hspace{7cm} {\large(d) 10 qubits}\\
{\includegraphics[width=0.48\textwidth]{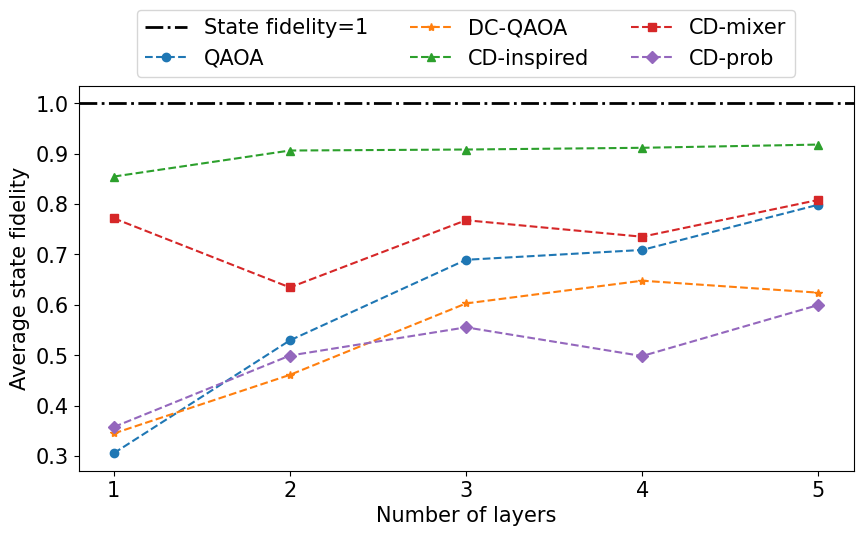}} \qquad{\includegraphics[width=0.48\textwidth]{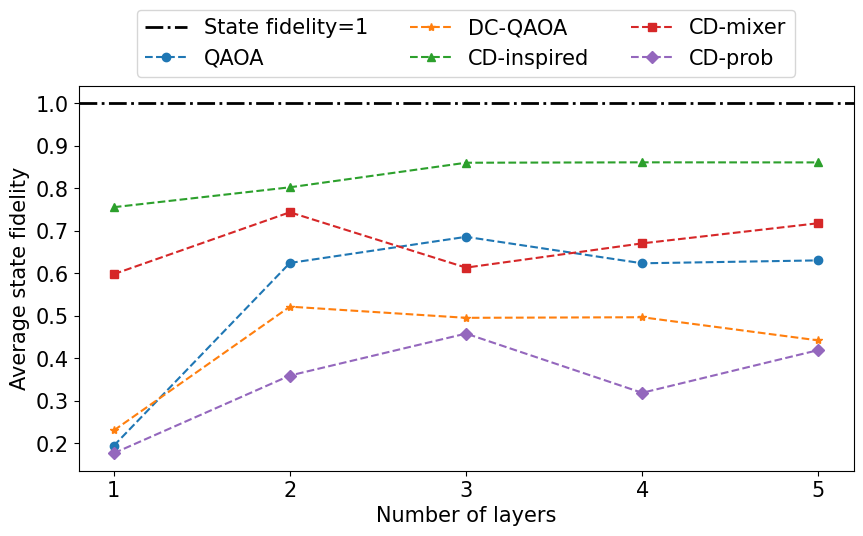}}
\end{tabular}

\caption{The average state fidelity between the $\h{H}_{Schwinger}$ ground state and the final ansatz state of the algorithms with an XY mixer for the Schwinger model against the number of layers for 4 qubits in (a), 6 qubits in (b), 8 qubits in (c), and 10 qubits in (d). The model parameters are $m = 0.35$, $w = 0.6$, $J = 5/12$ and $\theta = 0$.}
\label{sf_XY_mixer}

\end{figure*}

\begin{figure*}[!]
\centering

\begin{tabular}[b]{c}
\hspace{0.5cm}{\large(a) 4 qubits} \hspace{7cm} {\large(b) 6 qubits}\\
{\includegraphics[width=0.48\textwidth]{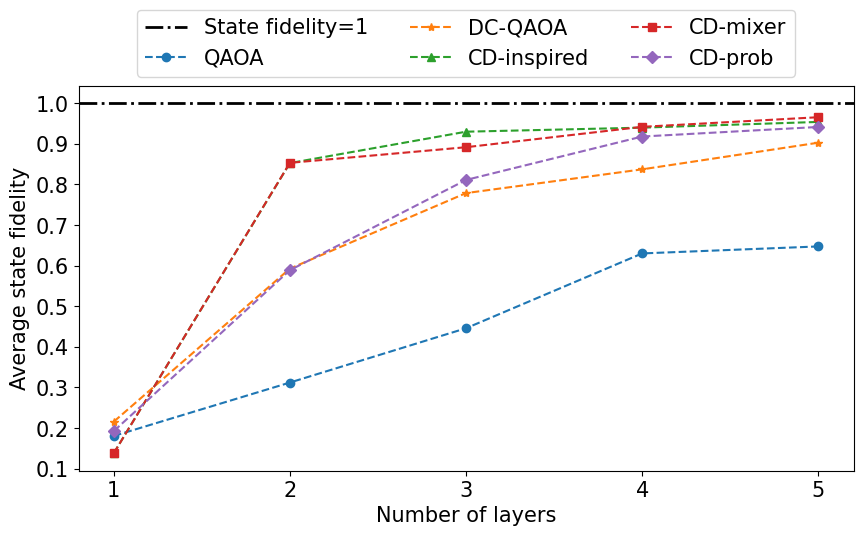}} \qquad{\includegraphics[width=0.48\textwidth]{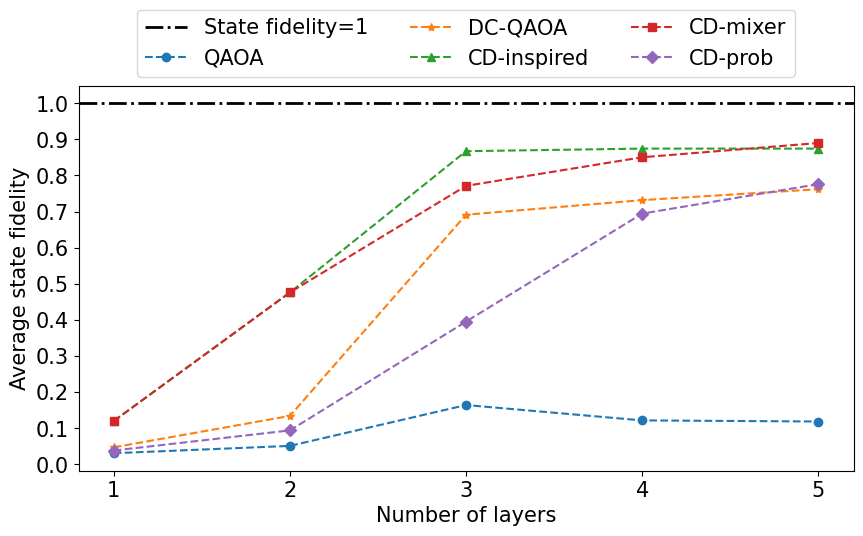}}
\end{tabular}

\begin{tabular}[b]{c}
\hspace{0.5cm}{\large(c) 8 qubits} \hspace{7cm} {\large(d) 10 qubits}\\
{\includegraphics[width=0.48\textwidth]{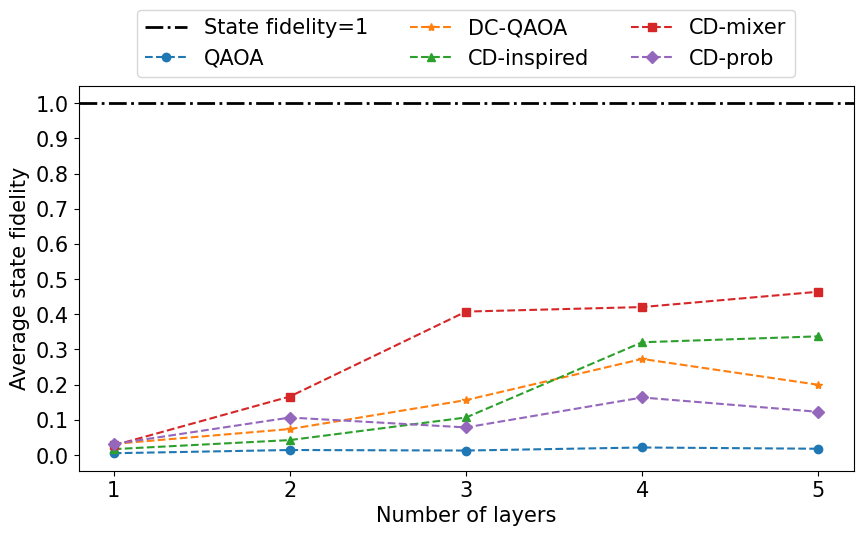}} \qquad{\includegraphics[width=0.48\textwidth]{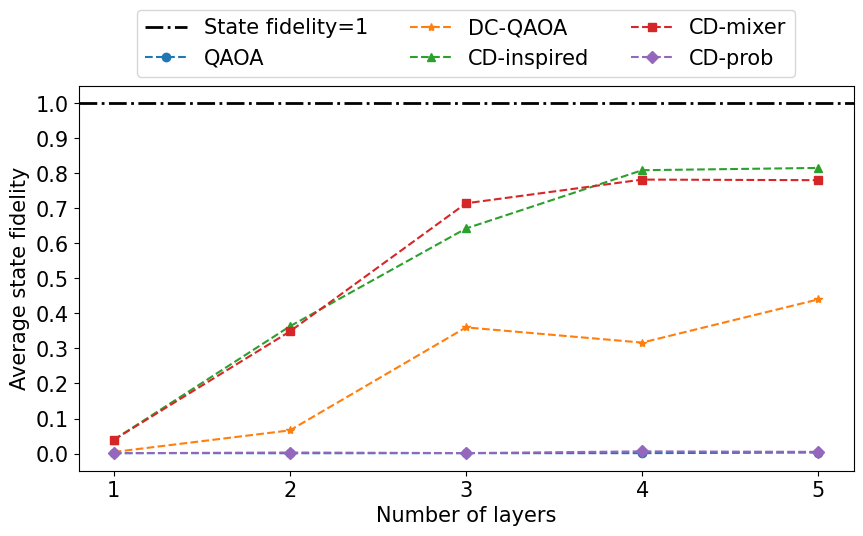}}
\end{tabular}

\caption{The average state fidelity between the $\h{H}_{Schwinger}$ ground state and the final ansatz state of the algorithms with an X mixer against the number of layers for and 4 qubits in (a), 6 qubits in (b), 8 qubits in (c), and 10 qubits in (d). The model parameters are $m = 0.35$, $w = 0.6$, $J = 5/12$ and $\theta = 0$.}

\label{sf_X_mixer}

\end{figure*}

The overall results presented in Figure \ref{XY_mixer} showcase how the CD-inspired variant performs best, especially at lower layers, hence making it a viable choice to be run at lower circuit depths. In fact, it is the only CD algorithm to consistently outperform QAOA for this mixer. Concerning the results obtained for the X mixer, presented in Figure \ref{X_mixer}, we observe that the CD-inspired and CD-mixer variants perform best, with them having a similar performance. In this case, all the CD algorithms consistently outperform QAOA. For a single layer, CD-inspired and CD-mixer yields a less pronounced improvement than with an XY mixer. However, for two or more layers, both the CD-inspired and the CD mixer achieve high approximation ratios, substantially outperforming QAOA.\\

Moreover, the experiments presented in Figures \ref{XY_mixer} and \ref{X_mixer} showcase how CD-inspired and CD-mixer scale well across the qubits tested, with them showing consistent results as the number of qubits is increased. An exception to this is the 8 qubit X mixer case where they achieve relatively lower approximation ratios when compared to the results concerning the other number of qubits. We believe that this discrepancy in the performance is an artefact of the gradient descent method when applied to a presumably more difficult optimization problem than others, with the optimizer struggling to converge and possibly getting stuck in other lower energy states. Concerning the other algorithms, broadly speaking, we observe that QAOA scales favourably with the XY mixer, whereas its performance with the X mixer declines as the number of qubits increases. Moreover, we observe that both DC-QAOA and CD-prob scale poorly with the XY mixer. Finally, we highlight the experimental observation that when using the the X mixer, increasing the qubit count yields a modest performance decline for DC-QAOA and a markedly greater deterioration for CD-prob.\\

On the other hand, the behaviour of DC-QAOA and CD-prob could have been expected as variational algorithms generally perform worse as the number of qubits increases. This is a consequence of the optimization landscape becoming more complicated due to the presence of many local minima and barren plateaus \cite{McClean_2018}. QAOA with an XY mixer, and CD-inspired and CD-mixer for both mixers, while employing the parameter fixing strategy, seem to avoid this issue, at least up to 10 qubits. Overall, we also observe that even for the cases where certain algorithms perform well, their approximation ratios tend to plateau to a point below 1 as the layers increase. This may also be due to local minima which are close to the global minima or from limitations of the classical optimisation procedure.\\

Concerning the choice of the mixer, we observe that the XY mixer performs better than the X mixer, especially for lower layers. This is most pronounced with QAOA, which performs the worse out of the algorithms with an X mixer and the second best with an XY mixer. As mentioned in Section \ref{Sec_3B_mixers}, this is expected as the XY mixer is more suitable for the Schwinger Hamiltonian compared to the X mixer. Despite QAOA performing poorly with an X mixer, the CD algorithms are often able to give significant improvements for higher layers, which indicates that QAOA performing well is not necessary for the CD algorithms to perform well. With an XY mixer the CD algorithms, with the exception of CD-inspired, generally do not perform better than QAOA, which indicates that these algorithms may not give an improvement if QAOA is already performing well.\\

\begin{table*}[!]
\resizebox{2\columnwidth}{!}{%
\begin{tabular}{||c|c||cccc||}
\hhline{|======|}
\multirow{2}{*}{Algorithm}   & \multirow{2}{*}{Mixer} & \multicolumn{4}{c||}{Fixed CD term set}                                                                                                             \\ \hhline{||~|~||====|} 
                             &                        & \multicolumn{1}{c||}{4 qubits}             & \multicolumn{1}{c|}{6 qubits}           & \multicolumn{1}{c||}{8 qubits}           & 10 qubits          \\ \hhline{|======|}
\multirow{2}{*}{DC-QAOA}     & XY                     & \multicolumn{1}{c||}{XY, YX, YX, XZY, YX}  & \multicolumn{1}{c||}{YX, XY, YX, YX, XY} & \multicolumn{1}{c||}{YX, XY, XY, XY, YX} & YX, XY, YX, YX, XY \\ \cline{2-6} 
                             & X                      & \multicolumn{1}{c||}{YX, ZY, XY, Y, YX}    & \multicolumn{1}{c||}{XY, Y, YX, YX, XY}  & \multicolumn{1}{c||}{YX, YX, Y, YX, Y}  & XY, Y, Y, Y, Y     \\ \hhline{|======|}
\multirow{2}{*}{CD-inspired} & XY                     & \multicolumn{1}{c||}{XY, XYZ, XY, XY, XY}  & \multicolumn{1}{c||}{XY, YX, XY, XY, YX} & \multicolumn{1}{c||}{XY, XY, YX, XY, YX} & XY, YX, XY, XY, XY \\ \cline{2-6} 
                             & X                      & \multicolumn{1}{c||}{YX, ZY, YX, YX, ZY}   & \multicolumn{1}{c||}{YX, ZY, YX, Y, Y}   & \multicolumn{1}{c||}{YX, XY, Y, XY, ZY}   & XY, Y, ZY, ZY, ZY  \\ \hhline{|======|}
\multirow{2}{*}{CD-mixer}    & XY                     & \multicolumn{1}{c||}{YX, XY, YX, XY, YZX}   & \multicolumn{1}{c||}{XY, XY, YX, YX, YX} & \multicolumn{1}{c||}{XY, YX, XY, YX, YX} & YX, XY, YX, YX, YX \\ \cline{2-6} 
                             & X                      & \multicolumn{1}{c||}{YX, ZY, YX, ZY, ZY}   & \multicolumn{1}{c||}{YX, ZY, Y, XY, ZY}  & \multicolumn{1}{c||}{YX, YZ, XY, Y, Y}   & XY, Y, ZY, Y, Y    \\ \hhline{|======|}
\multirow{2}{*}{CD-prob}     & XY                     & \multicolumn{1}{c||}{XY, ZXY, XY, ZXY, YX} & \multicolumn{1}{c||}{XY, XY, YX, YX, XY} & \multicolumn{1}{c||}{XY, YX, XY, YX, YX} & YX, XY, YX, YX, XY \\ \cline{2-6} 
                             & X                      & \multicolumn{1}{c||}{YX, ZY, XY, YX, YX}   & \multicolumn{1}{c||}{XY, ZY, XY, XY, XY} & \multicolumn{1}{c||}{YX, Y, Y, Y, XY}    & ZY, Y, XY, Y, Y    \\ \hhline{|======|}
\end{tabular}}
\caption{The fixed CD terms operators found when running the algorithms after 5 layers for different number of qubits and different mixers.}
\label{Table:H_CD_sets}
\end{table*}

In order to make sure that minimizing the energy is leading us towards the $\h{H}_{Schwinger}$ ground state, we can look at the state fidelity. Figure \ref{sf_XY_mixer} shows the average state fidelity of the final ansatz states of the algorithms with an XY mixer, and the $\h{H}_{Schwinger}$ ground state using 1 to 5 layers and 4, 6, 8 and 10 qubits. The ansatz states were taken from the same runs used in Figure \ref{XY_mixer}. We also present, in Figure \ref{sf_X_mixer}, the results obtained using an X mixer. For the sake of completeness, we also include in Appendix \ref{append: best data} the plots corresponding to the best state fidelity found during the runs with the XY mixer.\\

Comparing the results in Figures \ref{sf_XY_mixer} and \ref{sf_X_mixer} with Figures \ref{XY_mixer} and \ref{X_mixer} respectively we observe that, in general, a higher average approximation ratio leads to a higher average state fidelity, with the state fidelity plots showing a similar behaviour to the approximation ratio plots. This validates the approach of energy minimization to prepare ground states. For the XY mixer, as with the approximation ratios, CD-inspired is able to maintain high average state fidelities for all layers and qubits. For the X mixer, CD-inspired and CD-mixer are able to achieve high average state fidelities for higher layers. An exception to this is, again, the 8 qubit case where, similarly to the corresponding approximation ratios, the average state fidelities are noticeably worse than the other qubits.\\

We also observe that, for 4 qubits, the average state fidelities have very similar values to the average approximate ratios. As the number of qubits increase, the average state fidelity becomes lower than its corresponding average approximation ratio, with this gap increasing with increasing qubits due to the presence of more lower energy states. This issue may become significant if we go to higher qubits, with high approximation ratios not necessarily leading to high state fidelities due to the optimizer reaching low energy states rather than the ground state, in which case, energy minimization may not be the most suitable method to prepare the ground state.\\

In Table \ref{Table:H_CD_sets} we report the fixed CD term operators found from the algorithms after five layers. We observe for the XY mixer, above 4 qubits, that the operators are exclusively XY and YX. Therefore, it seems it is not necessary to include the three-body terms in the operator pool $A_{XY}$ as the two-body terms are more favoured. On the contrary, the fixed CD terms from the X mixers are much more varied.\\

\begin{table*}[!]
\resizebox{2\columnwidth}{!}{
\begin{tabular}{||c|c||cccccccccccc||}
\hhline{|==============|}
\multirow{3}{*}{Algorithms}  & \multirow{3}{*}{Mixer} & \multicolumn{12}{c||}{Number of qubits}                                                                                                                                                                                                                                                                                                                                                    \\ \hhline{||~|~||============|} 
                             &                        & \multicolumn{3}{c||}{4}                                                                       & \multicolumn{3}{c||}{6}                                                                       & \multicolumn{3}{c||}{8}                                                                       & \multicolumn{3}{c||}{10}                                                                      \\ \cline{3-14} 
                             &                        & \multicolumn{1}{l|}{CX gates} & \multicolumn{1}{l|}{Operations} & \multicolumn{1}{l||}{Depth} & \multicolumn{1}{l|}{CX gates} & \multicolumn{1}{l|}{Operations} & \multicolumn{1}{l||}{Depth} & \multicolumn{1}{l|}{CX gates} & \multicolumn{1}{l|}{Operations} & \multicolumn{1}{l||}{Depth} & \multicolumn{1}{l||}{CX gates} & \multicolumn{1}{l|}{Operations} & \multicolumn{1}{l||}{Depth} \\ \hhline{|==============|}
\multirow{2}{*}{QAOA}        & XY                     & \multicolumn{1}{c|}{86}       & \multicolumn{1}{c|}{386}        & \multicolumn{1}{c||}{222}   & \multicolumn{1}{c|}{277}      & \multicolumn{1}{c|}{988}        & \multicolumn{1}{c||}{639}   & \multicolumn{1}{c|}{805}      & \multicolumn{1}{c|}{3188}       & \multicolumn{1}{c||}{2146}  & \multicolumn{1}{c|}{2151}     & \multicolumn{1}{c|}{9082}       & 7112                       \\ \cline{2-14} 
                             & X                      & \multicolumn{1}{c|}{45}       & \multicolumn{1}{c|}{207}        & \multicolumn{1}{c||}{113}   & \multicolumn{1}{c|}{145}      & \multicolumn{1}{c|}{480}        & \multicolumn{1}{c||}{252}   & \multicolumn{1}{c|}{342}      & \multicolumn{1}{c|}{865}        & \multicolumn{1}{c||}{436}   & \multicolumn{1}{c|}{481}      & \multicolumn{1}{c|}{1278}       & 468                        \\ \hhline{|==============|}
\multirow{2}{*}{DC-QAOA}     & XY                     & \multicolumn{1}{c|}{147}      & \multicolumn{1}{c|}{526}        & \multicolumn{1}{c||}{381}   & \multicolumn{1}{c|}{461}      & \multicolumn{1}{c|}{1360}       & \multicolumn{1}{c||}{849}   & \multicolumn{1}{c|}{1190}     & \multicolumn{1}{c|}{3692}       & \multicolumn{1}{c||}{2471}  & \multicolumn{1}{c|}{2897}     & \multicolumn{1}{c|}{10428}      & 7699                       \\ \cline{2-14} 
                             & X                      & \multicolumn{1}{c|}{103}      & \multicolumn{1}{c|}{329}        & \multicolumn{1}{c||}{236}   & \multicolumn{1}{c|}{267}      & \multicolumn{1}{c|}{689}        & \multicolumn{1}{c||}{397}   & \multicolumn{1}{c|}{578}      & \multicolumn{1}{c|}{1296}       & \multicolumn{1}{c||}{636}   & \multicolumn{1}{c|}{764}      & \multicolumn{1}{c|}{1687}       & 705                        \\ \hhline{|==============|}
\multirow{2}{*}{CD-inspired} & XY                     & \multicolumn{1}{c|}{67}       & \multicolumn{1}{c|}{206}        & \multicolumn{1}{c||}{155}   & \multicolumn{1}{c|}{257}      & \multicolumn{1}{c|}{652}        & \multicolumn{1}{c||}{487}   & \multicolumn{1}{c|}{806}      & \multicolumn{1}{c|}{2452}       & \multicolumn{1}{c||}{1861}  & \multicolumn{1}{c|}{2184}     & \multicolumn{1}{c|}{8344}       & 6844                       \\ \cline{2-14} 
                             & X                      & \multicolumn{1}{c|}{48}       & \multicolumn{1}{c|}{114}        & \multicolumn{1}{c||}{88}    & \multicolumn{1}{c|}{179}      & \multicolumn{1}{c|}{305}        & \multicolumn{1}{c||}{200}   & \multicolumn{1}{c|}{233}      & \multicolumn{1}{c|}{400}        & \multicolumn{1}{c||}{200}   & \multicolumn{1}{c|}{364}      & \multicolumn{1}{c|}{668}        & 246                        \\ \hhline{|==============|}
\multirow{2}{*}{CD-mixer}    & XY                     & \multicolumn{1}{c|}{86}       & \multicolumn{1}{c|}{318}        & \multicolumn{1}{c||}{228}   & \multicolumn{1}{c|}{309}      & \multicolumn{1}{c|}{960}        & \multicolumn{1}{c||}{620}   & \multicolumn{1}{c|}{842}      & \multicolumn{1}{c|}{2902}       & \multicolumn{1}{c||}{2058}  & \multicolumn{1}{c|}{2444}     & \multicolumn{1}{c|}{9955}       & 7311                       \\ \cline{2-14} 
                             & X                      & \multicolumn{1}{c|}{50}       & \multicolumn{1}{c|}{144}        & \multicolumn{1}{c||}{99}    & \multicolumn{1}{c|}{110}      & \multicolumn{1}{c|}{286}        & \multicolumn{1}{c||}{149}   & \multicolumn{1}{c|}{352}      & \multicolumn{1}{c|}{704}        & \multicolumn{1}{c||}{341}   & \multicolumn{1}{c|}{364}      & \multicolumn{1}{c|}{737}        & 272                        \\ \hhline{|==============|}
\multirow{2}{*}{CD-prob}     & XY                     & \multicolumn{1}{c|}{118}      & \multicolumn{1}{c|}{417}        & \multicolumn{1}{c||}{302}   & \multicolumn{1}{c|}{427}      & \multicolumn{1}{c|}{1180}       & \multicolumn{1}{c||}{770}   & \multicolumn{1}{c|}{1128}     & \multicolumn{1}{c|}{3499}       & \multicolumn{1}{c||}{2375}  & \multicolumn{1}{c|}{2651}     & \multicolumn{1}{c|}{10023}      & 7466                       \\ \cline{2-14} 
                             & X                      & \multicolumn{1}{c|}{103}      & \multicolumn{1}{c|}{321}        & \multicolumn{1}{c||}{230}   & \multicolumn{1}{c|}{349}      & \multicolumn{1}{c|}{788}        & \multicolumn{1}{c||}{480}   & \multicolumn{1}{c|}{492}      & \multicolumn{1}{c|}{1090}       & \multicolumn{1}{c||}{603}   & \multicolumn{1}{c|}{961}      & \multicolumn{1}{c|}{1947}       & 849                        \\ \hhline{|==============|}
\end{tabular}}
\caption{The number of CX (controlled-not) gates, number of operations, and the depth of the circuits of the best ansatz states from the algorithms when transpiled through the fake IBM backend FakeAlgiers for 3 layers, 4, 6, 8 and 10 qubits, and with X and XY mixers. All the circuits were transpiled identically, with the optimization level set to 3 and the same transpiler seed.}
\label{Resource_estimate}
\end{table*}

To estimate the resource cost of the final ansatz states produced by each algorithm, the number of CX (controlled-not) gates, total number of operations, and circuit depth of the 3 layer ansatz state circuits were evaluated after transpiling the circuits onto the fake IBM backend FakeAlgiers. This was done using the fixed CD terms and the best-performing parameters (i.e., those yielding the lowest expectation value) identified from the runs shown in Figures \ref{XY_mixer} and \ref{X_mixer}. These results are shown in Table \ref{Resource_estimate}. FakeAlgiers is a fake 24 qubit backend with a native gate set of [CX, Id, RZ, SX, X] along with delay, reset and measure operations. It mimics real IBM quantum systems using snapshots of those systems. In the table, the number of CX gates is included as a measure of the resource cost because these two-qubit gates are generally more susceptible to noise and more resource intensive to implement than single-qubit gates. The remaining operations are made up of single qubit gates and reset operations. Note that this is not the resource cost of the full algorithms, which would require repeated measurements of the circuit to get expectation values and reruns of the circuit with updated parameters from the optimizer.\\

The data in table \ref{Resource_estimate} displays that the CD-inspired variant has the lowest number of operations and lowest depth for all qubits and for either mixer, which is expected as CD-inspired has the simplest ansatz state. An exception to this is the run with 6 qubits and an X mixer, where CD-mixer has the lowest number of operation and depth (as well as number of CX gates), which is caused by the fixed CD terms in CD-mixer being simpler than the ones in CD-inspired. Additionally, CD-inspired has a similar number of CX gates compared to QAOA, apart from for 8 and 10 qubits with an X mixer where QAOA has around a hundred more CX gates. Regarding the other algorithms, generally, CD-mixer has a similar number of operations and depth to QAOA for the XY mixer, and a lower amount for the X mixer, while DC-QAOA and CD-prob have the higher resource costs for all qubits and mixers. Furthermore, we observe that using an XY mixer can lead to a significant increase in the resource cost compared to an X mixer, which is due to the XY mixer being two-bodied and having a more complicated ground state. The data in Table \ref{Resource_estimate} is depended on the backend we use to transpile the circuits and the transpiler setting we choose, though we note that, in general, CD-inspired should produce states with a reduced number of operations and depth compared to the states produced by the other algorithms.

\begin{figure*}[!]

\centering

\includegraphics[width=0.8\textwidth]{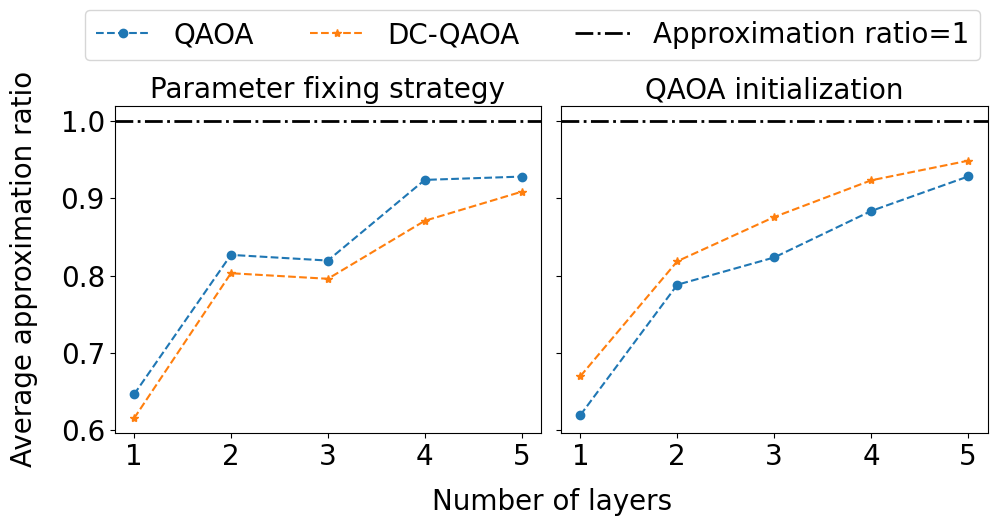}

\caption{The average approximation ratios for QAOA and DC-QAOA with an XY mixer for the Schwinger model against the number of layers for 6 qubits. On the left the variational parameters have been initialized using the parameter fixing strategy, and on the right they have been initialized using the QAOA initialization. The model parameters are $m = 0.35$, $w = 0.6$, $J = 5/12$ and $\theta = 0$.}

\label{Initialization comparison}
\end{figure*}

Overall, these results are somewhat surprising, as one would expect DC-QAOA to perform best due to having the most variational parameters and the most expressive ansatz state. However, it seems that the optimization may be having a considerable impact on the results. The additional variational parameters in DC-QAOA makes the optimization landscape of its expectation value more complicated, and even through the same paths towards the ground state that the other algorithms find exist in the landscape, the optimizer struggles to find them. The simpler landscapes of CD-inspired and CD-mixer allows the optimizer to find paths to the ground state more efficiently.\\

To investigate this further, in addition to the parameter fixing strategy that was described in Section \ref{subsec: Parameter fixing strategy} and employed as the parameter initialization for the results in Figures \ref{XY_mixer} and \ref{X_mixer}, we also considered a different initialization for QAOA and DC-QAOA. For this procedure, QAOA was ran as before using the parameter fixing strategy, but after each run its optimized parameters were passed as the $\boldsymbol{\gamma}$ and $\boldsymbol{\beta}$ parameters for DC-QAOA and the $\boldsymbol{\alpha}$ parameters were set to zero. This starts the optimizer in the QAOA minima meaning its performance should be similar or better than QAOA. We refer to this initialization as the QAOA initialization.\\ 

Figure \ref{Initialization comparison} presents a comparison between the two different parameter initializations for QAOA and DC-QAOA with an XY mixer and using 6 qubits. Note that for DC-QAOA we still employ the CD term fixing, described in Section \ref{subsec: CD term fixing}, for both initializations, and the approximation ratios are the average approximation ratios obtained for the fixed CD term from the pool for that layer. Unlike with the parameter fixing strategy, where QAOA performs better than DC-QAOA, with the QAOA initialization DC-QAOA outperforms QAOA as expected. This indicates that QAOA outperforming DC-QAOA is an artefact of the parameter initialization, as we can obtain the opposite result with a different initialization. Additional results from using the QAOA initialization for 4, 8 and 10 qubits are presented in Appendix \ref{append: QAOA init}.

\section{Conclusion} \label{Sec_5_conclusion}

In this paper, we have looked at applying the CD algorithms DC-QAOA, CD-inspired, CD-mixer and CD-prob to the prepare the ground state of the Schwinger model and compared their performances with the standard QAOA. Additionally, we also utilized the parameter fixing strategy and CD term fixing in these algorithms to achieve improved results. We found that when running the algorithms with an X mixer, a case where QAOA does not perform well for this model, the CD algorithms had an improved performance over QAOA, and for DC-QAOA, CD-inspired and CD-mixer, this improvement was significant for higher layers. When running the algorithms with an XY mixer, a case where QAOA does perform well for this model, CD-inspired was able to consistently outperform QAOA with an improved performance over its X mixer results. CD-inspired and CD-mixer also showed good scaling, with them performing similarly across all numbers of qubits that were tested. This makes CD-inspired with both mixers and CD-mixer with the X mixers very viable algorithms for this problem.\\

Additionally, the final ansatz state circuits produced by CD-inspired for both mixers and CD-mixer for the X mixer had a lower number of operations and depth than QAOA with the respective mixer, making them easier to run on an actual quantum computer. A possible disadvantage is that CD-inspired and CD-mixer need to be ran more times than QAOA due to the CD term fixing method that was employed however, they are able to attain better results at lower layers which may mitigate some of this cost. Overall, the prospect of using counterdiabatic algorithms to prepare the ground state of the Schwinger model and other gauge theories looks very promising.\\

\section{Acknowledgements}

We would like to thank Alex Tomlinson for helpful advice and discussions. This work was supported by an Engineering and Physical Sciences Research Council Doctoral Training Partnership Research Studentship in quantum technology. Research of BC at the University of Southampton has been supported by the following research grants - STFC (Grant no. ST/X000583/1), STFC (Grant no. ST/W006251/1), and EPSRC (Grant no. EP/W032635/1). We acknowledge the use of the IRIDIS High Performance Computing Facility, and associated support services at the University of Southampton, in the completion of this work.

\bibliography{Literature}

\renewcommand{\arraystretch}{2}

\begin{figure*}[!ht]
\centering

\begin{tabular}[b]{c}
\hspace{0.5cm}{\large(a) 4 qubits} \hspace{7cm} {\large(b) 6 qubits}\\
{\includegraphics[width=0.48\textwidth]{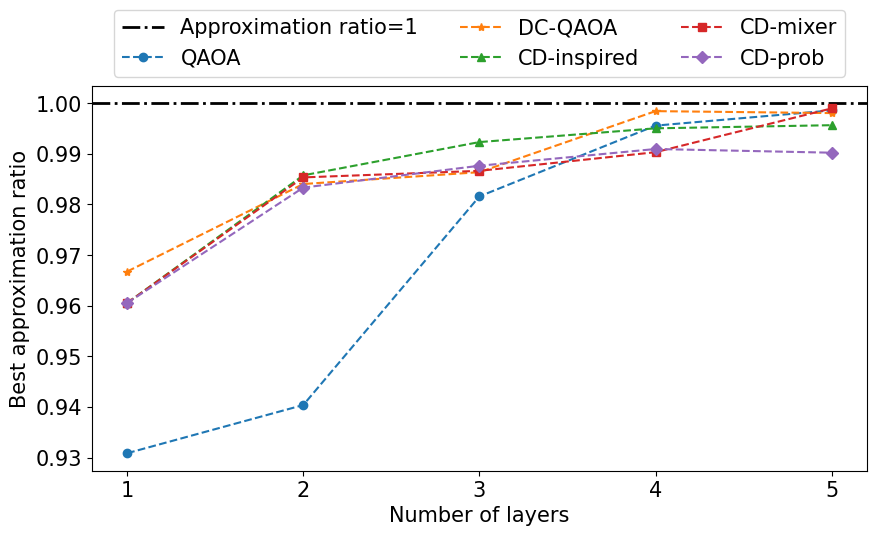}} \qquad{\includegraphics[width=0.48\textwidth]{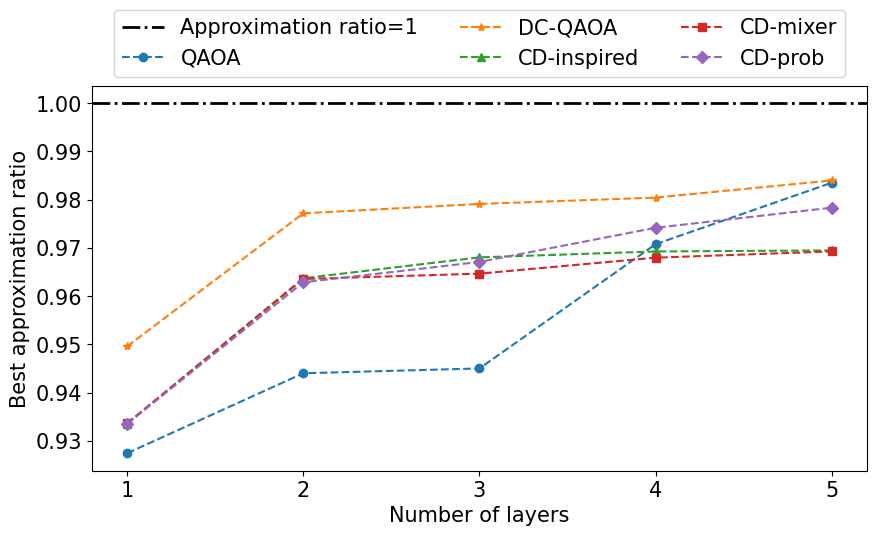}}
\end{tabular}

\begin{tabular}[b]{c}
\hspace{0.5cm}{\large(c) 8 qubits} \hspace{7cm} {\large(d) 10 qubits}\\
{\includegraphics[width=0.48\textwidth]{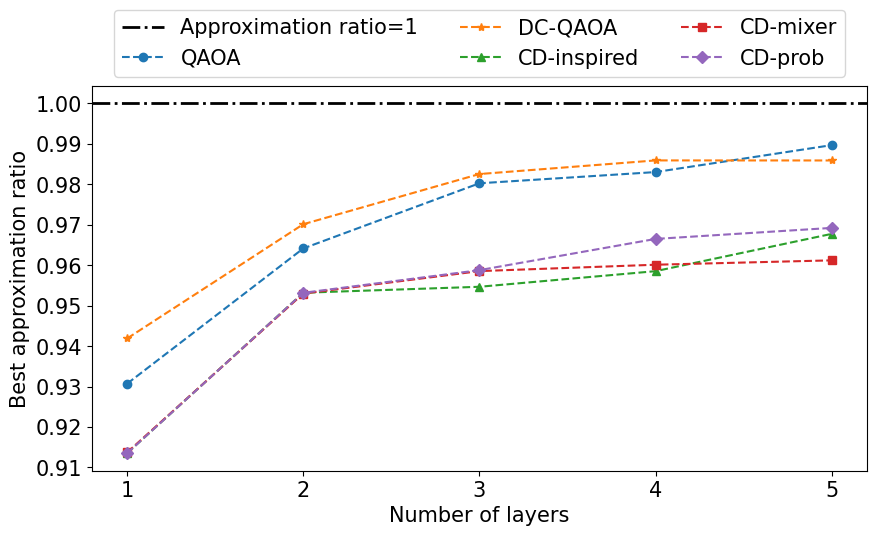}} \qquad{\includegraphics[width=0.48\textwidth]{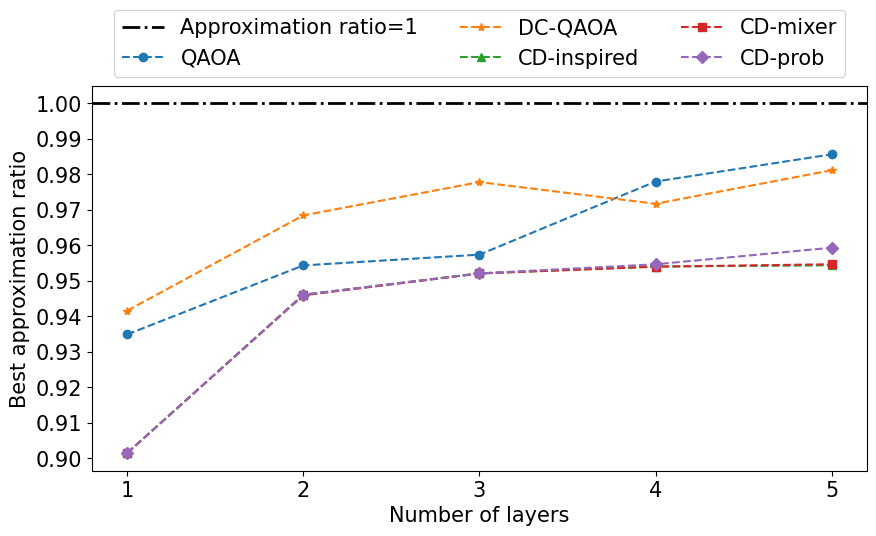}}
\end{tabular}

\caption{The best approximation ratios of the algorithms with an XY mixer for the Schwinger model, against the number of layers for 4 qubits in (a), 6 qubits in (b), 8 qubits in (c), and 10 qubits in (d). The model parameters are $m = 0.35$, $w = 0.6$, $J = 5/12$ and $\theta = 0$.}
\label{XY_mixer_best_AP}

\end{figure*}

\renewcommand{\arraystretch}{2}

\begin{figure*}[!]
\centering

\begin{tabular}[b]{c}
\hspace{0.5cm}{\large(a) 4 qubits} \hspace{7cm} {\large(b) 6 qubits}\\
{\includegraphics[width=0.48\textwidth]{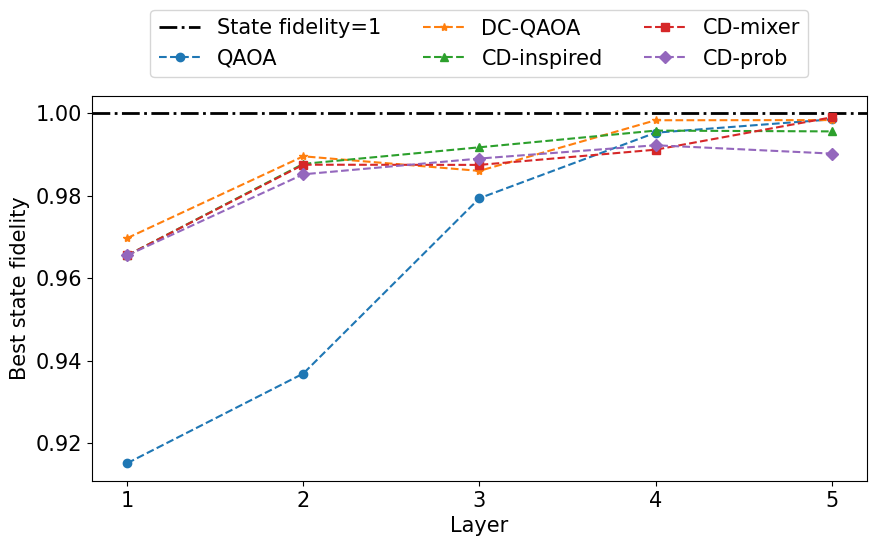}} \qquad{\includegraphics[width=0.48\textwidth]{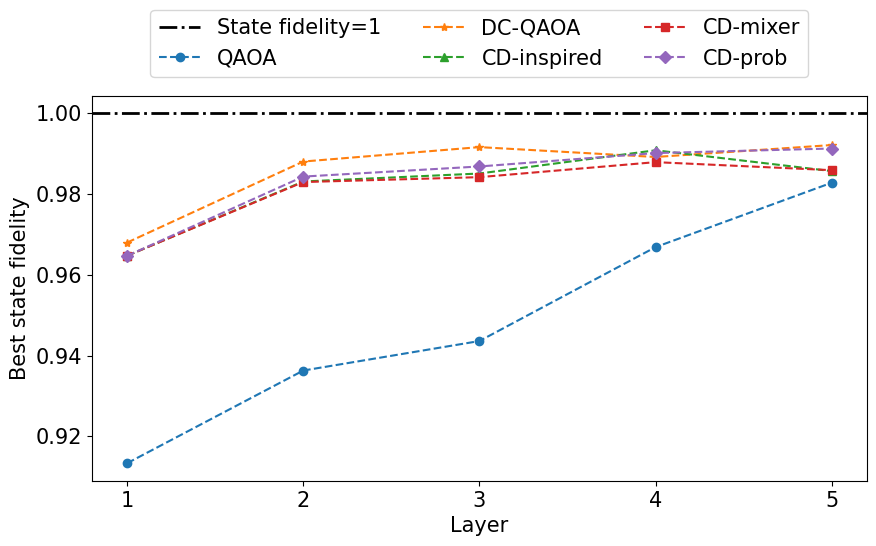}}
\end{tabular}

\begin{tabular}[b]{c}
\hspace{0.5cm}{\large(c) 8 qubits} \hspace{7cm} {\large(d) 10 qubits}\\
{\includegraphics[width=0.48\textwidth]{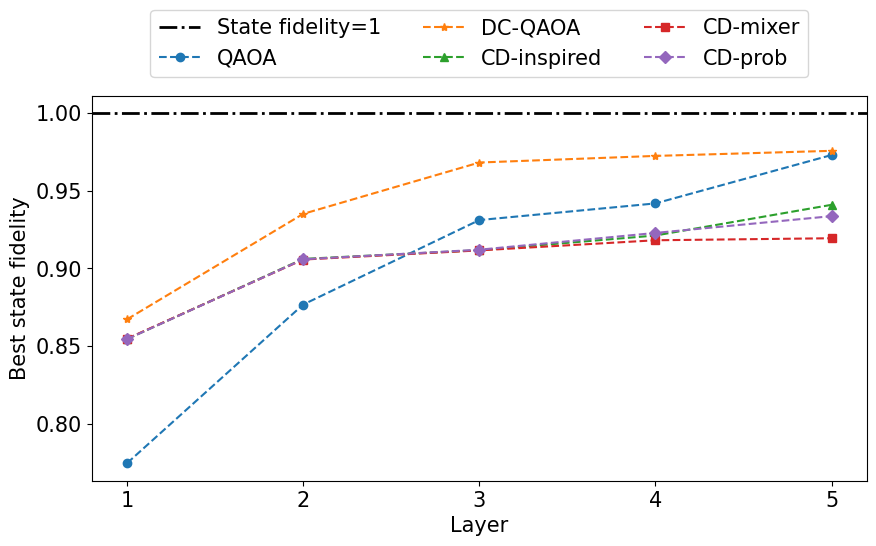}} \qquad{\includegraphics[width=0.48\textwidth]{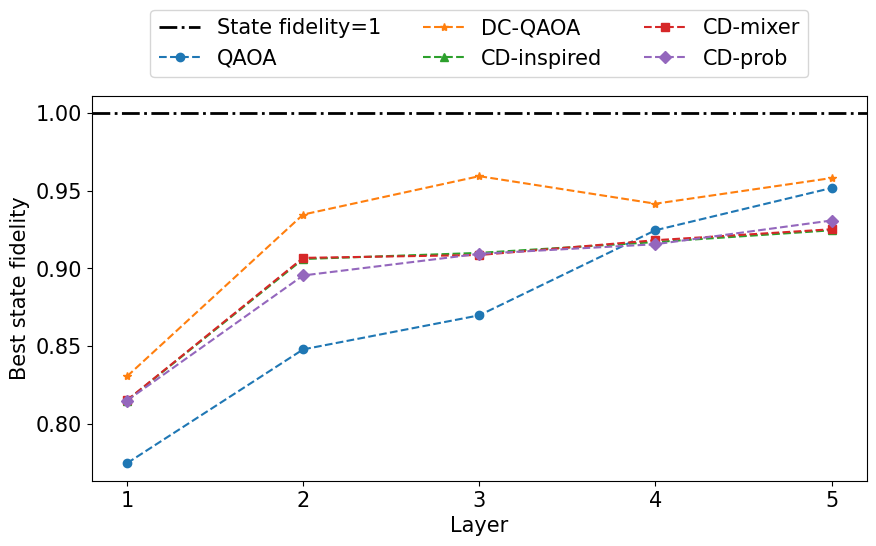}}
\end{tabular}

\caption{The best state fidelity between the $\h{H}_{Schwinger}$ ground state and the final ansatz state of the algorithms with an X mixer against the number of layers for 4 qubits in (a), 6 qubits in (b), 8 qubits in (c), and 10 qubits in (d). The model parameters are $m = 0.35$, $w = 0.6$, $J = 5/12$ and $\theta = 0$.}
\label{XY_mixer_best_SF}

\end{figure*}

\renewcommand{\arraystretch}{2}

\begin{figure*}[!]
\centering

\begin{tabular}[b]{c}
\hspace{0.5cm}{\large(a) 4 qubits} \hspace{7cm} {\large(b) 6 qubits}\\
{\includegraphics[width=0.48\textwidth]{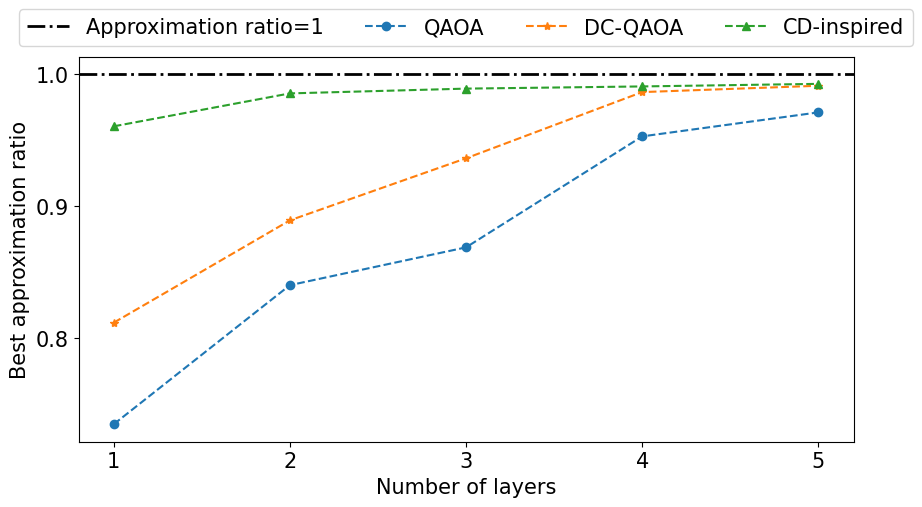}} \qquad{\includegraphics[width=0.48\textwidth]{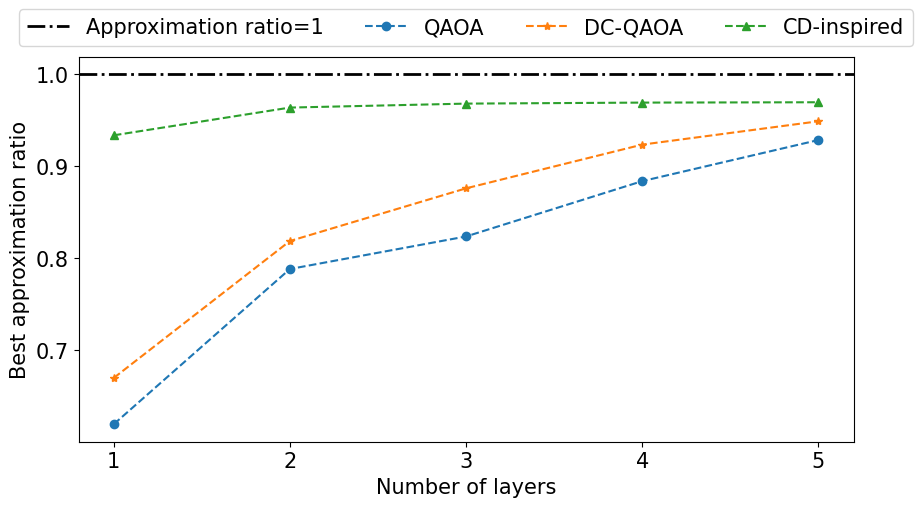}}
\end{tabular}

\begin{tabular}[b]{c}
\hspace{0.5cm}{\large(c) 8 qubits} \hspace{7cm} {\large(d) 10 qubits}\\
{\includegraphics[width=0.48\textwidth]{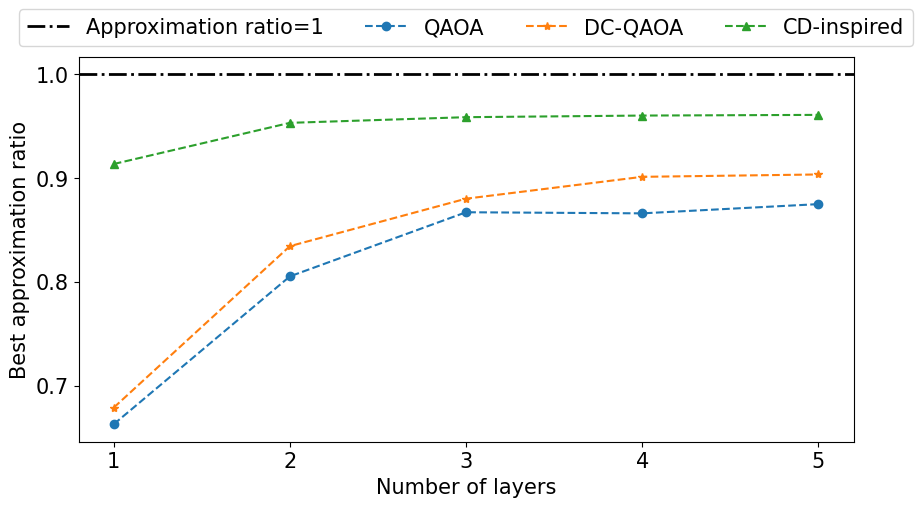}} \qquad{\includegraphics[width=0.48\textwidth]{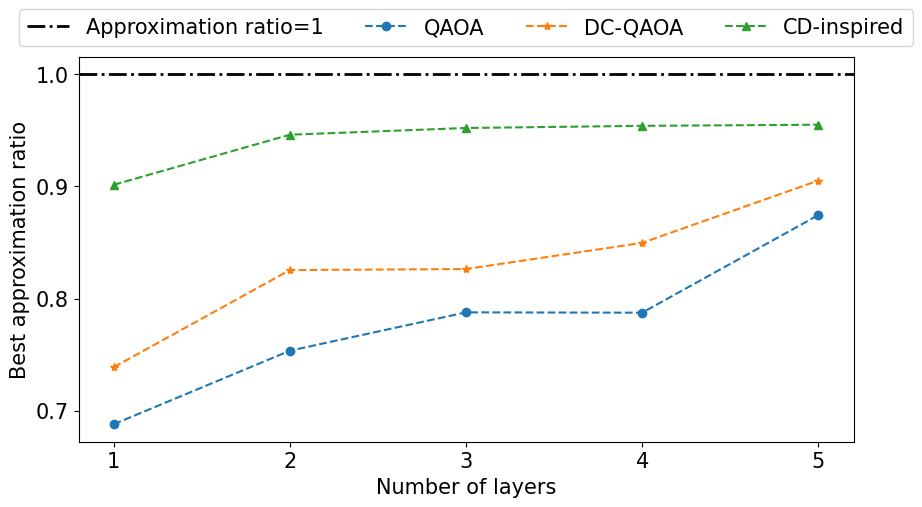}}
\end{tabular}

\caption{The average approximation ratios of the QAOA, DC-QAOA, and CD-inspired with an XY mixer for the Schwinger model using the QAOA initialization for the initial variational parameters against the number of layers for 4 qubits in (a), 6 qubits in (b), 8 qubits in (c), and 10 qubits in (d). The model parameters are $m = 0.35$, $w = 0.6$, $J = 5/12$ and $\theta = 0$.}
\label{XY_mixer_QAOA_init}

\end{figure*}

\appendix

\section{Best approximation ratios and state fidelity for the XY mixer} \label{append: best data}

In section \ref{Sec_4_results_discussion} we evaluated the performance of the algorithms by conducting multiple runs with different initial variational parameter values, and calculating the average approximation ratio and the average state fidelity of the final ansatz state with the $\h{H}_{Schwinger}$ ground state. However, another metric we can look at is the best approximation ratio and best state fidelity. This means instead of taking averages of the results we only look at the highest approximation ratio and state fidelity that was found during the runs. This can be a useful measurement as our goal in using these algorithms is to output an approximate ground state for the Schwinger model.\\ 

For each layer, QAOA was ran 20 times, while the CD algorithms were ran 20 times for each term in its $\h{A}_{DC}$ operator pool. This may give the CD algorithms an advantage since they were ran more times. We only show the results for the algorithms with an XY mixer as this yielded a better performance compared to the X mixer. The best results are found from the same set of results used for the previous plots and tables with the XY mixer.\\

Figure \ref{XY_mixer_best_AP} shows the best approximation ratios of the algorithms with an XY mixer for the Schwinger model using 1 to 5 layers and 4, 6, 8 and 10 qubits. We observe that all the algorithms maintain high approximation ratios, and the difference between performances of the algorithms is less pronounced compared to the average approximation ratios in Figure \ref{XY_mixer}. Furthermore, for 4 qubits, the CD algorithms all perform similarly and better than QAOA below 4 layers, while for 6 qubits DC-QAOA performs the best, and for 8 and 10 qubits, DC-QAOA and QAOA perform best with them having a similar performance. Overall, these results indicate that QAOA and DC-QAOA are able to achieve a better performance with higher approximation ratios compared to the other algorithms. However, CD-inspired is able to more reliably achieve good results as it has a higher average approximation ratio compared to the other algorithms.\\

Figure \ref{XY_mixer_best_SF} shows the best state fidelity of the final ansatz states of the algorithms with an XY mixer and the $\h{H}_{Schwinger}$ ground state using 1 to 5 layers and 4, 6, 8 and 10 qubits. Similar to the best approximation ratios in Figure \ref{XY_mixer_best_AP}, we see that all the algorithms are able to achieve very high state fidelities, and the difference between the performances of the algorithms is less pronounced compared to the average state fidelities in Figure \ref{sf_XY_mixer}. Moreover, the state fidelities are lower for 8 and 10 qubits compared to 4 and 6 qubits, similar to what was observed with the average state fidelities. Comparing the plots with their corresponding best approximation ratios plots in Figure \ref{XY_mixer_best_AP}, we note that the 4 qubit plot has a similar pattern, while for 6 qubit there is a bigger gap between QAOA and the CD algorithms. Additionally, for 8 and 10 qubits, there are some cases of CD-inspired, CD-mixer and CD-prob performing better than QAOA despite QAOA having a higher best approximation ratio. This indicates that even though QAOA may produce states with energies closer to the Schwinger ground state compared to the other algorithms, this does not always correspond to a better Schwinger ground state. Note that the runs with the best approximation ratio did not necessarily have the best state fidelity, though, in general, runs with higher approximation ratios tended to have higher state fidelities.\\

\section{QAOA initialization for the XY mixer} \label{append: QAOA init}

In section \ref{Sec_4_results_discussion} we briefly looked at a different way to initialize the variational parameters for QAOA and DC-QAOA to investigate why QAOA was outperforming DC-QAOA. We referred to this as the QAOA initialization, where QAOA was ran first using the parameter fixing strategy, described in Section \ref{subsec: Parameter fixing strategy}, and after every run its optimized parameters were passed into DC-QAOA as the $\boldsymbol{\gamma}$ and $\boldsymbol{\beta}$ parameters with the $\boldsymbol{\alpha}$ parameters set to zero. In Figure \ref{XY_mixer_QAOA_init} we show the average approximation ratios of QAOA and DC-QAOA with an XY mixer using this initialization using 1 to 5 layers and 4, 6, 8 and 10 qubits. Additionally, we have also included the results for CD-inspired with an XY mixer. Since CD-inspired does not have any $\boldsymbol{\gamma}$ and $\boldsymbol{\beta}$ parameters, it is ran with all its $\boldsymbol{\alpha}$ parameters set to zero. Since the AerSimulator measures expectation values exactly, when we run CD-inspired with the same $\hat{A}_{DC}$ terms and same initial variational parameters values, we get the same results for each run, so the average approximation ratio is the same as the actual approximation ratio. Note that for DC-QAOA and CD-inspired, we still employ the CD term fixing, described in Section \ref{subsec: CD term fixing}, for this initialization, and the approximation ratios are the average approximation ratios obtained for the fixed CD term from the pool for that layer\\

As with the 6 qubit case in Figure \ref{Initialization comparison}, we see that DC-QAOA is performing better than QAOA, but it is not a significant improvement. Moreover, CD-inspired is still able to outperform both algorithms, which indicates that the optimization for CD-inspired is easier, allowing the optimizer to find lower minima despite not having the QAOA minima in its optimization landscape. Note that for the 10 qubit case, the learning rate of the optimizer was change from the default of 0.01 to 0.001, as with the higher value we found the optimizer was starting in QAOA minima but immediately moving to a higher energy region in the optimization landscape.

\end{document}